\documentclass[twocolumn,prb,aps,longbibliography]{revtex4-2}

\usepackage{dcolumn}
\usepackage{amsmath}
\usepackage{amsfonts}
\usepackage{amssymb}
\usepackage{calligra}
\usepackage{bm}
\usepackage{graphicx}
\usepackage{float}
\usepackage{subfigure}
\usepackage{natbib}
\usepackage{physics}
\usepackage[]{hyperref}
\usepackage{wasysym}
\begin{document}

\title{Probing Electron Localization and Delocalization in the Selective Long-Range Tight-Binding Model}

\author{Mohammad Pouranvari}
\email{m.pouranvari@umz.ac.ir}
\affiliation{Department of Solid-State Physics, Faculty of Science,
University of Mazandaran, Babolsar, Iran.}

\date{\today}

\begin{abstract}
In this study, we perform a detailed investigation into the interplay between disorder-induced electron localization and long-range hopping amplitudes within the Selective Long-Range Tight-Binding Model (SLRTB). Through numerical simulations, we analyze the electronic properties of the system, with a focus on the participation ratio (PR), entanglement entropy (EE), energy spectrum, and the ratio of level spacings ($r_n$). Our results reveal a marked distinction between negative and positive long-range hopping amplitudes, manifesting in different electronic behaviors and transitions. Notably, we carry out a finite-size scaling analysis, identifying the critical point and exponents that characterize the system's behavior near the transition. The investigation highlights the role of gapless regions in shaping the system's PR, $r_n$, and EE, and the influence of disorder on these properties. The SLRTB model proves to be an effective framework for understanding the effects of disorder and long-range hopping on electron dynamics, offering valuable insights into localization and delocalization phenomena.
\end{abstract}

\maketitle

\section{Introduction} \label{sec:introduction}

In the realm of condensed matter physics, Anderson localization stands as a cornerstone, illuminating the profound consequences of disorder on the electronic behavior of materials\cite{PhysRev.109.1492}. This phenomenon, first articulated by Philip W. Anderson in 1958, elucidates how the presence of disorder, in the form of random potential variations or impurities, can lead to a remarkable quantum mechanical effect — the localization of electronic wave functions. Such localization signifies that electrons become confined to small, spatially isolated regions within the material, hindering their ability to propagate freely\cite{lagendijk2009fifty, semeghini2015measurement,
	PhysRevB.5.2931}.

Central to Anderson localization is the notion that when disorder exceeds a critical strength, electronic states that would otherwise extend throughout the material, allowing for electronic transport, become highly localized\cite{markos2006numerical,
	doi:10.1080/13642819308215292, RevModPhys.80.1355,
	PhysRevLett.82.4062, mirlin1996transition}. This localization manifests as a striking absence of electronic mobility and a transition from extended states to decaying wave functions. The implications of Anderson localization reverberate across diverse domains of physics, ranging from condensed matter physics to solid-state chemistry\cite{doi:10.1021/j100061a002, doi:10.1021/acsnano.9b02399, https://doi.org/10.1002/aelm.201700050, Chen2014}.

Notably, the ramifications of Anderson localization are far-reaching, impacting the electronic properties and transport characteristics of a wide array of materials, including semiconductors\cite{PhysRevB.107.174204,PhysRevLett.75.3914}, insulators\cite{PhysRevB.92.125143, PhysRevLett.114.216601,PhysRevLett.94.056404, PhysRevB.92.085410}, and even disordered electronic systems\cite{khomchenko2023influence,Yan:23}. This phenomenon underpins our understanding of insulating behavior in disordered materials and has profound implications for the development of electronic devices, quantum computing, and quantum technologies.

Our research is motivated by the intricate interplay of disorder, long-range hopping, and selectivity in electronic systems. These three factors collectively shape the behavior of electrons in condensed matter physics.

Firstly, disorder is well-known to induce localization in electronic systems\cite{thakkar2023effect, schwartz2007transport,lahini2008anderson, PhysRevB.76.214204}. Disorder disrupts the regular crystal lattice, leading to a localization of electronic states. Understanding this effect is crucial as it has significant implications in various fields, including semiconductor physics and materials science.

On the other hand, long-range hopping interactions have the opposite effect, promoting delocalization of electronic states\cite{doi:10.1126/science.1232296, perez2019interplay,celardo2016shielding,cantin2018effect, chattaraj2016effects, nag2019many, chavez2021disorder, klein1993localization}. These interactions extend over relatively long distances, allowing electrons to move freely across the material. This behavior is of great interest for its potential applications in quantum materials and device engineering.

However, our motivation extends further due to the presence of selectivity in our system. The selective nature of the long-range interactions introduces a unique dimension to the problem, and we seek to unravel how selectivity, disorder, and long-range hopping combine to influence electronic behavior.

By investigating this intricate interplay through numerical simulations of the SLRTB model, we aim to provide insights into the behavior of energy levels, participation ratios, ratios of level spacings, and entanglement entropy. Our research addresses fundamental questions about how these factors collectively govern electron behavior.Ultimately, our motivation lies in advancing our understanding of electron localization and delocalization in complex, disordered materials with selective long-range interactions.

The remainder of this paper is organized as follows. In Section \ref{model}, we provide an overview of the theoretical framework and model parameters governing our investigation, with a focus on the impact of on-site energies on electron localization and delocalization and the balancing of positive and negative on-site energies. Section \ref{measures} introduces the measures, including the participation ratio (PR), ratio of level spacings ($r_n$), and entanglement entropy (EE), which serve as crucial indicators of electron behavior in the context of the Selective Long-Range Tight-Binding (LRTB) model. In Section \ref{results}, we present the results of our numerical calculations, offering insights into the behavior of energy levels, PR, $r_n$, and EE.  Finally, we conclude our study in Section \ref{conclusion}, summarizing key observations and their implications, and we provide insights into future directions in the field of electron localization and delocalization.

\section{Theoretical Framework and Model Parameters}\label{model}
In this section, we establish the theoretical foundation of our investigation and introduce the pivotal parameters that define SLRTB model. The model's fundamental components, including the Hamiltonian structure, on-site energies, and hopping amplitudes, are presented, setting the stage for a comprehensive exploration of electron localization and delocalization within the context of disordered systems.

The central element of our investigation is encapsulated in the Hamiltonian presented as:

\begin{equation}
	H = \sum_{i=1}^{N} \epsilon_i c^{\dagger}_i c_{i} + \sum_{i,j} t_{ij} c^{\dagger}_i c_{j} + \text{h.c.}
\end{equation}

In this equation, $N$ denotes the system size (the number of lattice sites), and $c_i$ ($c^{\dagger}_i$) represents the annihilation (creation) fermion operator at site $i$. The on-site energies, $\epsilon_i$, are uniformly distributed within the range $[-W/2, W/2]$, where $W$ signifies the disorder strength. The hopping amplitudes, $t_{ij}$, depend on the on-site energies: $t_{ij} = t$ if both sites $i$ and $j$ exhibit negative on-site energies; otherwise, $t_{ij} = 0$.

In this study, we set the Fermi energy $E_F = 0$ at temperature $T = 0$. Since the system is at zero temperature, this implies that the chemical potential is also $\mu = 0$. We focus on the behavior of the system near $E_F = 0$ to explore the localization-delocalization transition in disordered systems.

Anderson localization, a phenomenon of paramount significance in the realm of condensed matter physics, is intricately influenced by the interplay of two fundamental factors within the context of SLRTB model. These factors are as follows:

Disorder and Randomness: Intrinsic to our model is the presence of disorder, primarily manifesting through the stochastic distribution of on-site energies ($\epsilon_i$) across the lattice. The spatial heterogeneity of on-site energies introduces a quenched random potential landscape that scatters electrons. As a result, electron wavefunctions become localized in specific regions of the lattice. This localization arises due to the multiple scattering events that electrons undergo as they navigate through the disordered medium. The consequence of this disorder-induced localization is a characteristic of the Anderson localization phenomenon. It is worth noting that even infinitesimal disorder can lead to the localization of electronic states, as exemplified in the one-dimensional conventional Anderson model.

Long-Range Hopping Mechanism: Distinguishing our selective long-range Hamiltonian from the conventional Anderson model is the introduction of a controlled long-range hopping mechanism. In our model, electrons are allowed to hop between arbitrary sites $i$ and $j$ only if both sites exhibit negative on-site energies, leading to a non-locality in the Hamiltonian. This long-range hopping mechanism introduces a crucial element that can counteract the localization tendencies induced by disorder. It provides a means for electrons to traverse the lattice over relatively long distances, potentially enabling them to avoid localization and remain delocalized. The role of long-range hopping in mitigating localization is a central focus of our investigation. It may lead to a competition between localization and delocalization depending on the specific values of $t$ and $W$.

The behavior of electrons in disordered systems is influenced by the interaction between \textit{on-site energies} ($\epsilon_i$) and \textit{hopping amplitudes} ($t_{ij}$). These factors collectively govern whether electrons become localized or delocalized.

On-site energies and localization: Both positive and negative on-site energies contribute to localization or delocalization depending on their \textit{relative values} to neighboring sites. If an electron at a given site has an on-site energy that is much lower than neighboring sites, it can become localized due to the formation of a potential well. Conversely, if the on-site energy at a site is much higher relative to its neighbors, it acts as a potential barrier, discouraging localization at that site.

Thus, \textit{localization} occurs not simply due to negative or positive on-site energies, but because of the \textit{relative energy differences} between adjacent sites. The balance between \textit{disorder} (random on-site energies) and the resulting \textit{energy landscape} creates regions where electrons may localize or delocalize, depending on the energy contrast across the lattice.

Interplay of hopping amplitudes and on-site energies:
The \textit{hopping amplitudes} ($t_{ij}$) play a crucial role in determining whether electrons can overcome energy barriers and delocalize across the lattice. The sign and magnitude of $t_{ij}$ influence how easily an electron can move between sites with different on-site energies.

For \textit{positive hopping amplitudes} ($t_{ij} > 0$), electrons tend to migrate from regions of higher energy to lower energy, facilitating delocalization. For \textit{negative hopping amplitudes} ($t_{ij} < 0$), the situation is more complex, as electrons can still move from higher energy sites to lower ones, but the overall dynamics are affected by the negative hopping terms.

In systems governed by \textit{Hamiltonian dynamics} without external perturbations (closed systems), the total energy is conserved. Therefore, electron localization or delocalization depends not on an absolute energy gain or loss, but on the \textit{relative energy differences} between adjacent sites. These energy differences shape how effectively electrons can hop between sites and overcome potential barriers.

\section{Quantitative Measures}\label{measures}
In this section, we introduce essential quantitative measures and observables that serve as crucial tools in our investigation. These measures include the Participation Ratio (PR), Ratio of Level Spacing ($r_n$), and Entanglement Entropy (EE). Their definitions and significance will be elucidated in preparation for their application and analysis in the subsequent section.

\subsection{Participation Ratio (PR)}

The Participation Ratio (PR) is a quantitative measure used to gauge the extent of electron localization within a quantum system\cite{murphy2011generalized,pruisken1985participation,calixto2015inverse}. It offers valuable insights into the distribution of electron wavefunctions across different sites in a lattice, providing a means to distinguish between localized and delocalized states.

\begin{equation}
	PR = \frac{1}{\sum_{i=1}^{N} |\psi_i|^4}.
\end{equation}

In scenarios where the PR approaches a value of 1, it indicates that electrons are highly localized within the system. Consider a hypothetical situation where an electron is confined to a single lattice site. In this case, the PR would be close to 1, signifying that the electron's wavefunction is concentrated at a single point. This represents a state of complete electron localization, akin to an electron 'trapped' within a specific region of the system. Conversely, when the PR tends towards a larger value (closer to the system size), it signifies a state of electron delocalization. Imagine the opposite scenario, where an electron's wavefunction spans the entire lattice, equally distributed across all sites. In such cases, the PR would approach the system size, indicating that the electron is not confined to a specific location but is spread out throughout the lattice. This corresponds to a state of complete electron delocalization, resembling the behavior of a free electron. In most real-world scenarios, the behavior lies between the extremes of complete localization and complete delocalization. The PR captures this intermediate behavior, reflecting the degree to which electrons are localized or delocalized within the system.

\subsection{Ratio of Level Spacing ($r_n$)}
The Ratio of Level Spacing ($r_n$) is a quantitative measure used to characterize the distribution of energy level spacings in a quantum system. It provides valuable insights into the system's behavior, particularly with regard to electron localization and delocalization\cite{PhysRevB.47.11487, PhysRevLett.123.025301, biroli2012differencelevelstatisticsergodicity, PhysRevB.97.125116}:

\begin{equation}
	r_n = \frac{\min(\Delta_n, \Delta_{n+1})}{\max(\Delta_n, \Delta_{n+1})},
\end{equation}
where $\Delta_n = E_{n+1}-E_n$ is difference between adjacent energy eigen-states for energy spectrum
$\{E\}$). In the delocalized phase, where electrons are spread out and mobile across the system, $r_n$ typically approaches a value of approximately $0.386$. This value indicates a statistically uniform distribution of energy level spacings, signifying the absence of strong localization effects. Conversely, in the localized phase where electrons are confined to specific regions of the lattice, $r_n$ tends to be larger, around $0.530$.  The values $ r_n \approx 0.386 $ and $ r_n \approx 0.530 $ are known from earlier works to characterize different phases\cite{PhysRevLett.110.084101, PhysRevB.75.155111}. This higher value indicates an irregular distribution of energy level spacings, reflecting the presence of localized states with larger level spacings and gaps between them.

\subsection{Entanglement Entropy (EE)}

We consider a many-body quantum system described by a pure ground state eigenstate $\ket{\Psi}$ at absolute zero temperature. In this context, the density matrix is defined as $\rho = \ket{\Psi}\bra{\Psi}$. The quantum system is divided into two distinct subsystems, denoted as $A$ and $B$. The reduced density matrix for each subsystem is obtained by tracing out the degrees of freedom associated with the other subsystem, yielding $\rho^{A} = \text{tr}_{B} (\rho)$.

The Block von Neumann entanglement entropy, which characterizes the quantum entanglement between the two subsystems\cite{PhysRev.47.777, RevModPhys.73.565,
	schrodinger_1935, Osterloh2002, RevModPhys.80.517,
	RevModPhys.81.865}., can be expressed as follows:
\begin{equation}\label{EE_definition}
	EE = -\text{tr}(\rho^{A}\ln{\rho^{A}}) = -\text{tr}(\rho^{B}\ln{\rho^{B}}).
\end{equation}

To compute the entanglement energies, we employ correlation functions as outlined in Ref. \cite{Klich_2006}. The eigenvalues $\{ \zeta \}$ of the correlation matrix for subsystem $A$,
\begin{equation}\label{correlation_matrix}
	C_{i,j} = \langle c_{i} ^{\dagger} c_{j} \rangle,
\end{equation}
(where $i$ and $j$ vary from $1$ to $N_A$, the subsystem size), are calculated. The block von Neumann entanglement entropy (EE) is given by:
\begin{equation}\label{EE_calculation}
	EE = -\sum_{i=1} ^{N_A} \left[ \zeta_i \ln(\zeta_i) + (1-\zeta_i) \ln(1-\zeta_i) \right].
\end{equation}

Entanglement entropy (EE) plays a crucial role in distinguishing between localized and delocalized phases in quantum many-body systems. It is generally observed that EE tends to be smaller in the localized phase and larger in the delocalized phase. This discrepancy arises due to the contrasting nature of electron correlations in these phases. In the localized phase, wavefunctions are spatially confined to specific regions, limiting correlations between particles, which results in lower EE. Conversely, in the delocalized phase, wavefunctions are extended, allowing for stronger correlations between distant parts of the system, leading to larger EE values \cite{jia2008entanglement,andrade2014anderson,laflorencie2022entanglement,pastur2014area,PhysRevB.89.115104,Pouranvari2020,POURANVARI2023128908,PhysRevB.100.195109}.

This general behavior has been consistently observed across various models of non-interacting fermions, particularly in systems exhibiting metal-insulator transitions. In these systems, EE grows logarithmically or follows a power-law in the metallic (delocalized) phase, often violating the entanglement area law in one-dimensional systems, while no such violation is observed in the insulating (localized) phase \cite{PhysRevB.89.115104, https://doi.org/10.1155/2015/397630}.

This behavior is consistent with results from both one-dimensional and higher-dimensional systems. In gapped, local models, EE generally follows area laws, where EE saturates with increasing block size \cite{wolf2006violation, gioev2006entanglement}. In higher-dimensional systems, small violations of area laws have been observed in critical fermionic models, often manifesting as logarithmic corrections \cite{PhysRevA.74.022329, li2006scaling1, li2006scaling2}. In disordered free fermion systems in the Anderson model, EE obeys the area law in one, two, and three dimensions as long as the subsystem size is larger than the mean free path, particularly in the metallic phase where the mean free path is finite although the localization length is infinite \cite{https://doi.org/10.1155/2015/397630}.

Despite these exceptions, the general trend remains that EE is larger in delocalized phases due to the extended nature of wavefunctions and the increased quantum correlations they enable. This relationship between EE and wavefunction extension serves as a useful tool for identifying transitions between localized and delocalized phases in disordered systems \cite{PhysRevLett.98.220603}.

\section{Results}\label{results}

In this section, we present the key findings and outcomes of our investigation into electron localization and delocalization in a selective long-range Hamiltonian model. We delve into the intriguing interplay between disorder-induced localization and the mitigating effects of long-range hopping amplitudes, shedding light on the intricate quantum mechanical behavior of electrons. Through numerical simulations, we explore the influence of various parameters, such as disorder strength ($W$), and hopping amplitudes ($t$), on the system's behavior.

Understanding the energy spectrum of a quantum system is essential for unraveling its electronic properties and phase transitions. In SLRTB model, characterized by a controlled long-range hopping mechanism and randomly distributed on-site energies, the energy levels and gaps hold key insights into the behavior of electrons. These gaps, or lack thereof, play a pivotal role in determining the system's electronic behavior, specifically the distinction between localized and delocalized phases. Here, we present the intriguing patterns of energy levels and gaps within our model.

\begin{figure}
	\centering
	\begin{subfigure}{}%
		\includegraphics[width=0.23\textwidth]{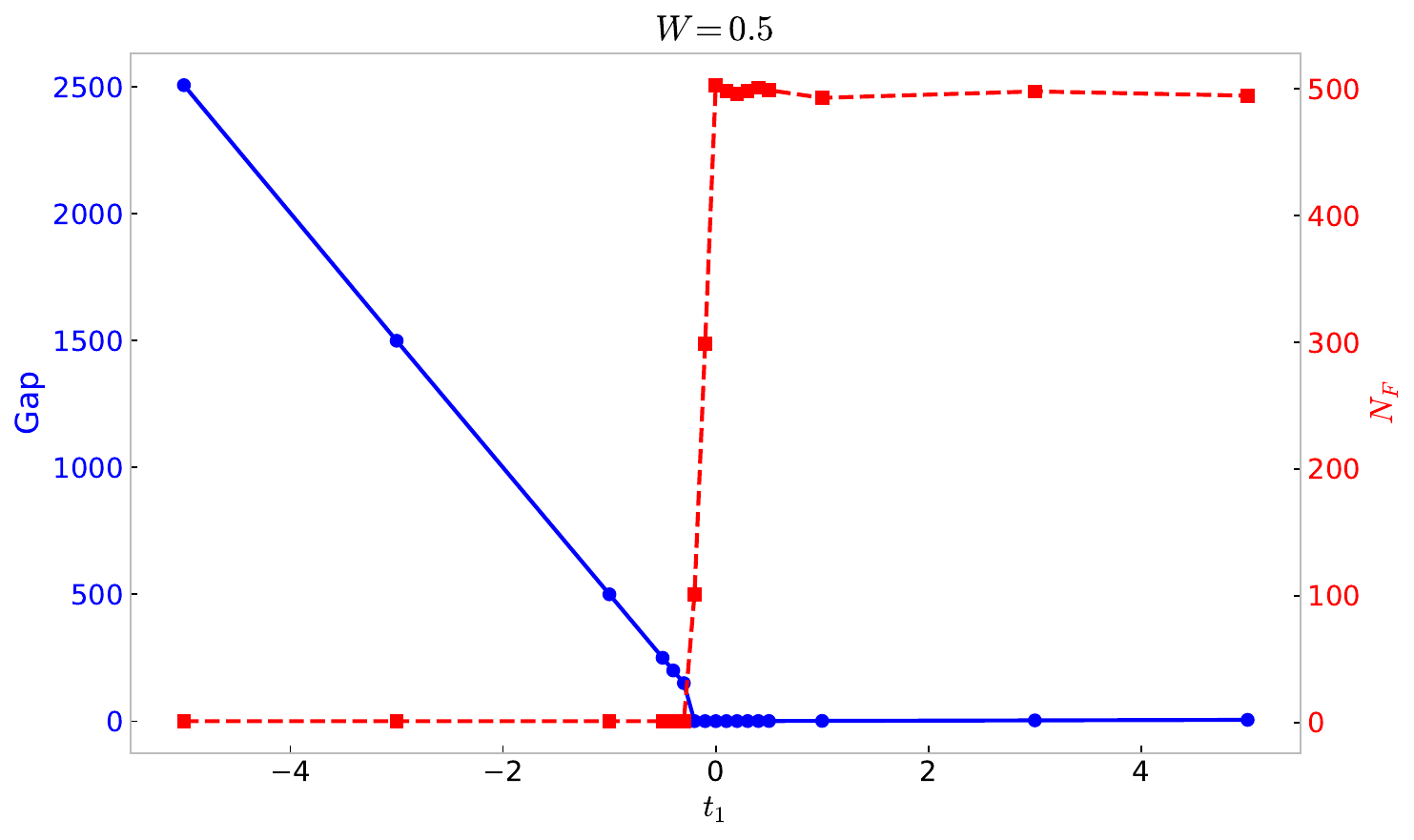}
	\end{subfigure}
	\begin{subfigure}{}%
		\includegraphics[width=0.23\textwidth]{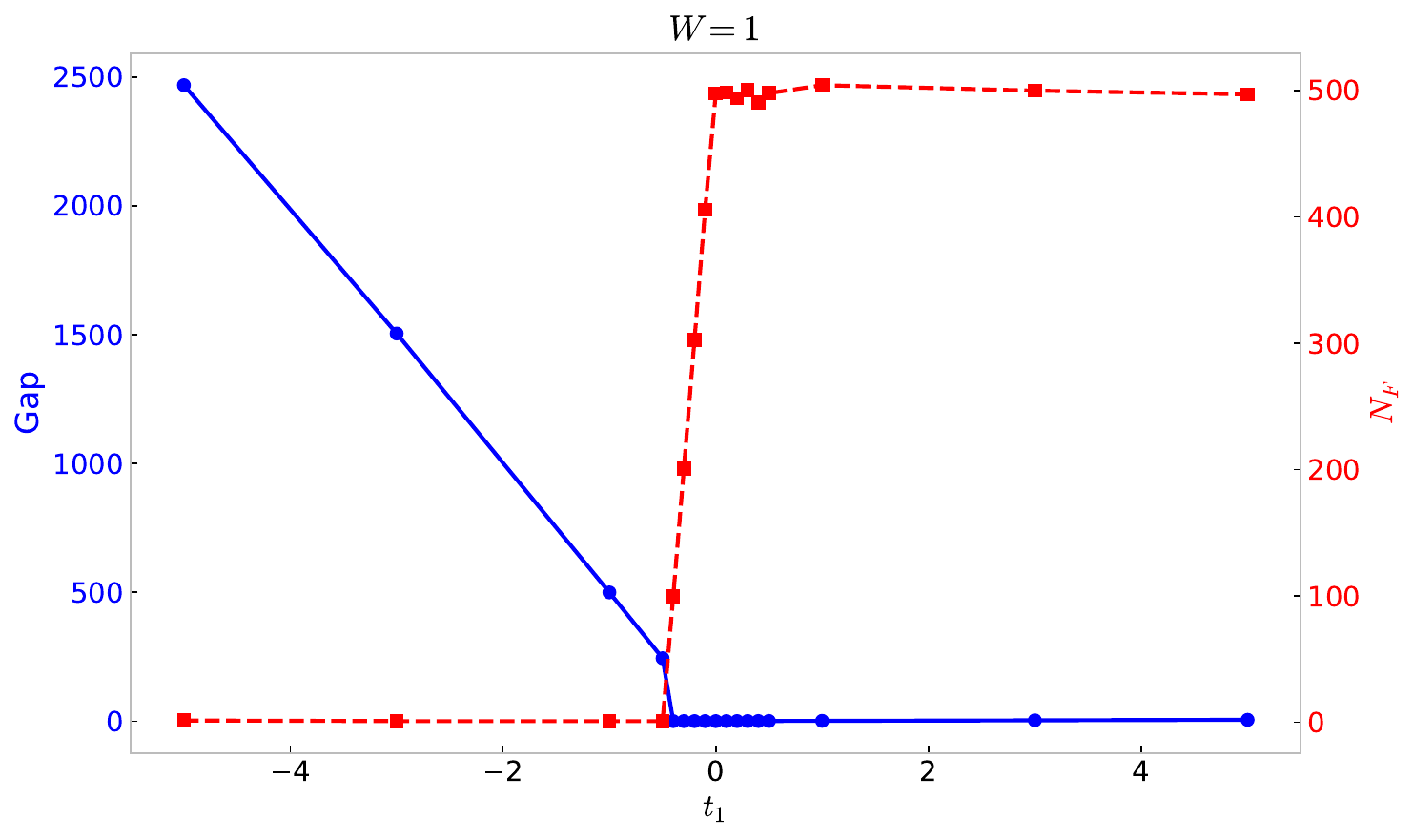}
	\end{subfigure}%
	\begin{subfigure}{}%
		\includegraphics[width=0.23\textwidth]{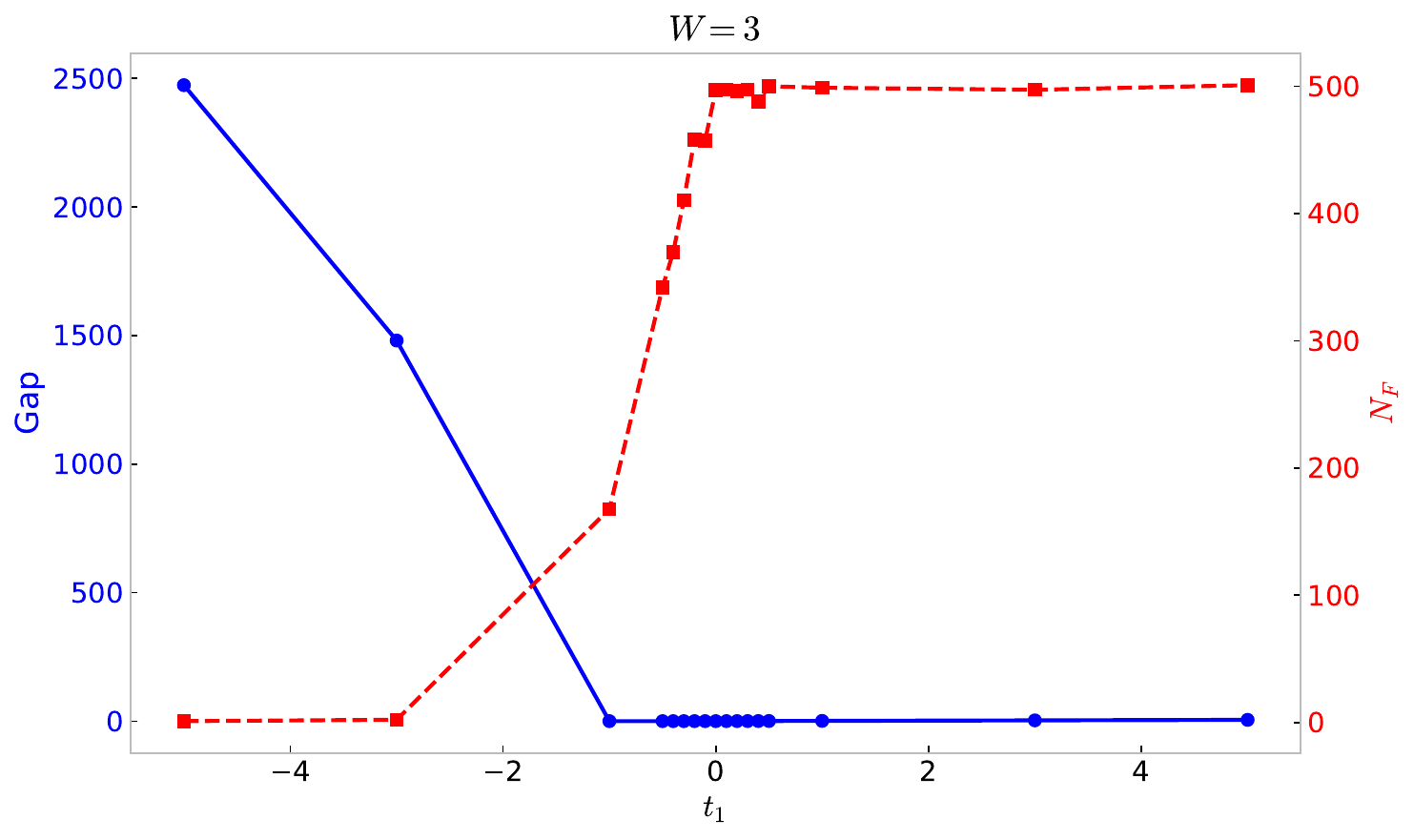}
	\end{subfigure}%
	\begin{subfigure}{}%
		\includegraphics[width=0.23\textwidth]{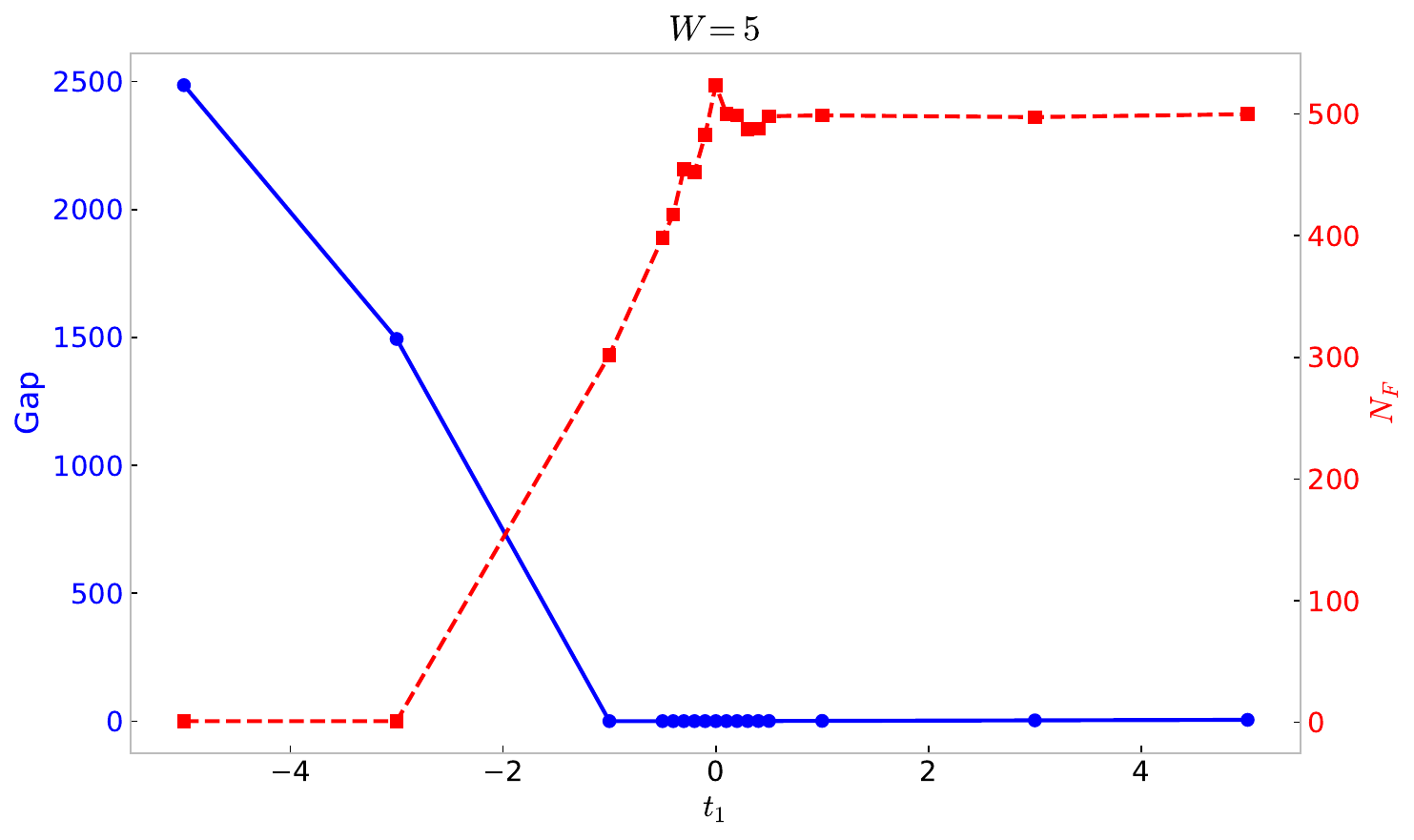}
	\end{subfigure}%
	\caption{Energy gap and number of filled states ($N_F$) as a function of the hopping amplitude $t$ for different disorder strengths $W$. The gap (blue line) and $N_F$ (red dashed line) are plotted for various values of $W$, showing how the energy gap opens for positive $t$, while gapless regions are present for negative $t$. As $W$ increases, the gapless region extends to larger negative values of $t$, and the number of filled states transitions from 1 to $N/2$ as $t$ increases. The results are obtained with $N=1000$ and sample averaging over $200$ realizations.\label{fig:energy_gaps}}
\end{figure}

In Figure \ref{fig:energy_gaps}, we present the behavior of the disorder averaged energy gaps and the disorder averaged number of filled states ($N_F$) as functions of the hopping amplitude parameter $t$ for different disorder strengths $W$ in the SLRTB model. The energy gap represents the difference between the highest filled energy level and the lowest unfilled energy level at the Fermi energy $E_F = 0$, while $N_F$ corresponds to the number of states below $E_F$.

Each plot shows the gap and $N_F$ for a specific disorder strength $W$. In general, for positive values of $t$, the system tends to be gapless, as indicated by a zero or near-zero energy gap. For negative values of $t$, the system becomes gapped, where a clear separation between filled and unfilled states emerges. The behavior of $N_F$ shows that, in general, the number of fermions is 1 for negative values of $t$, and it grows to $N/2$ as $t$ increases to positive values. However, both the energy gap and $N_F$ deviate from this trend in a region close to small negative values of $t$. As the disorder strength $W$ increases, this discrepancy becomes more pronounced and extends to larger negative $t$ values. This behavior reflects the influence of disorder on the system, where the gapless region shifts and broadens with increasing disorder.

\begin{figure}
	\centering
	\begin{subfigure}{}%
		\includegraphics[width=0.23\textwidth]{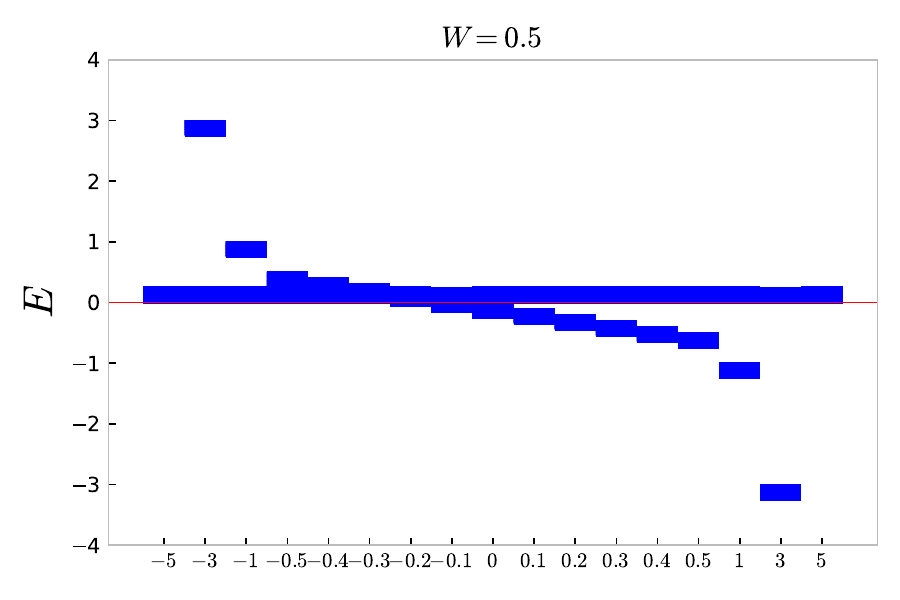}
	\end{subfigure}
	\begin{subfigure}{}%
		\includegraphics[width=0.23\textwidth]{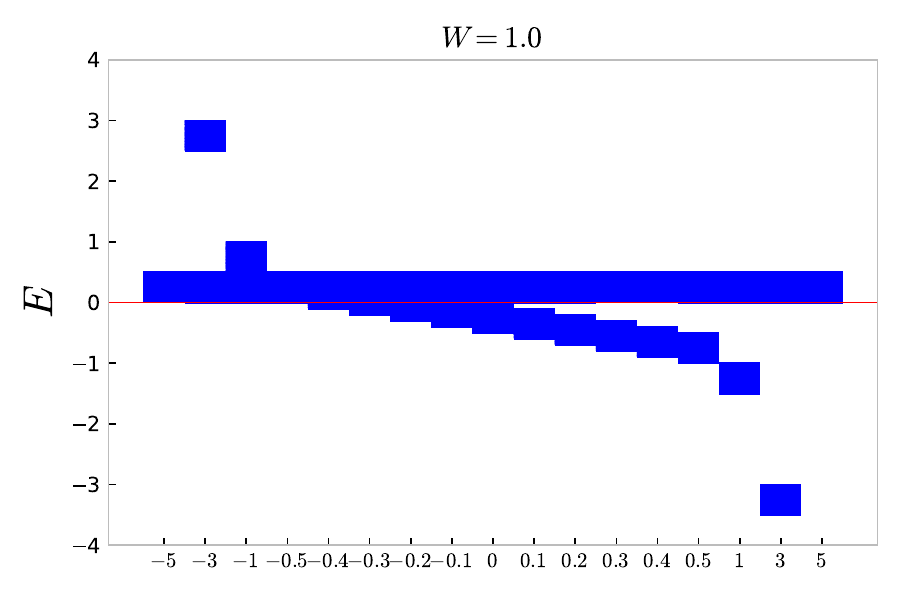}
	\end{subfigure}%
	\begin{subfigure}{}%
		\includegraphics[width=0.23\textwidth]{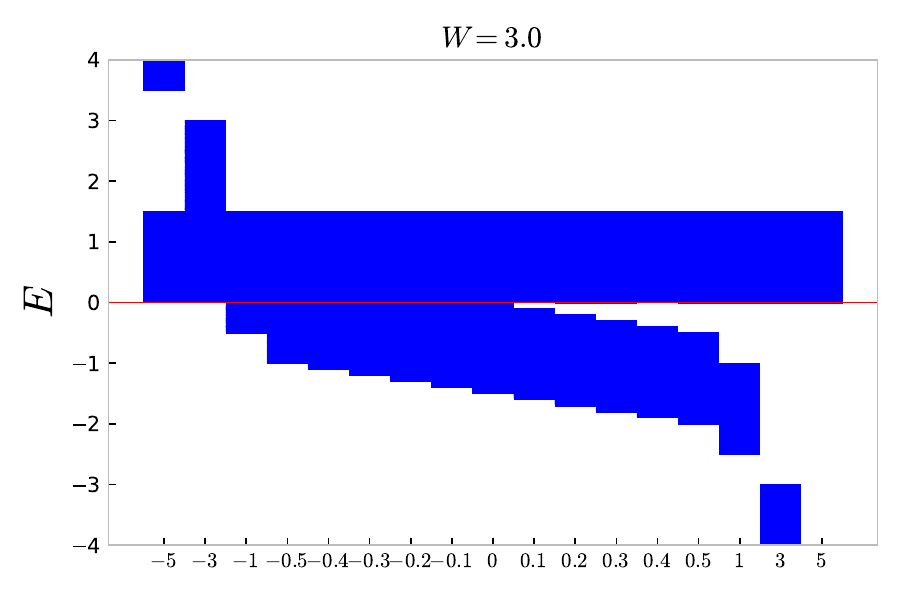}
	\end{subfigure}%
	\begin{subfigure}{}%
		\includegraphics[width=0.23\textwidth]{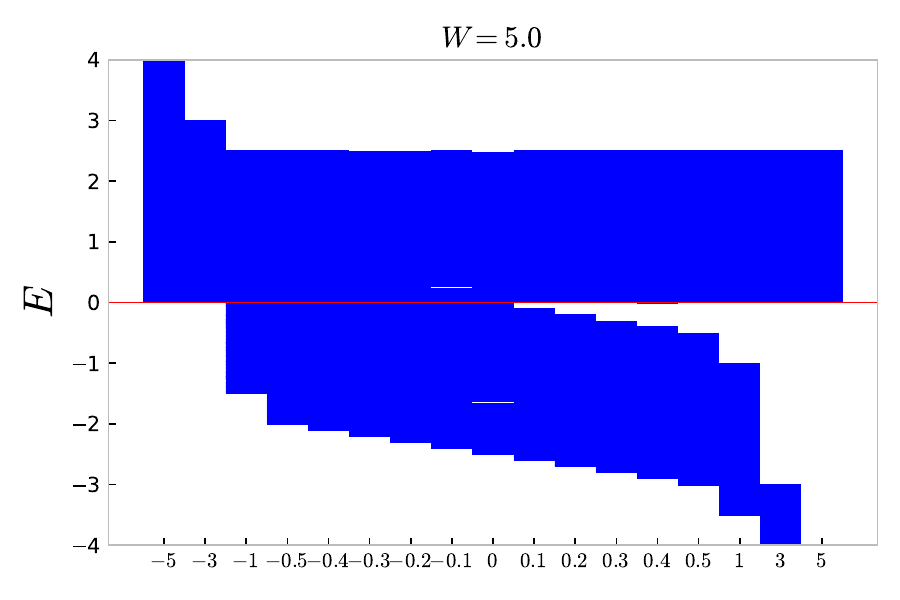}
	\end{subfigure}%
	\caption{Zoomed-in view of the energy gaps around the Fermi level ($E_F = 0$) for various disorder strengths $W$. Each plot shows energy levels within the range $[-4, 4]$ to provide a clearer illustration of the energy gap structure. Panel (a) corresponds to $ W = 0.5 $, Panel (b) to $ W = 1 $, Panel (c) to $ W = 3 $, and Panel (d) to $ W = 5 $. The zoomed plots are intended to highlight the gaps more distinctly and facilitate understanding of the energy level distribution and gap characteristics. \label{fig:energy_gaps_zoom}}
\end{figure}

To provide a clearer view of the energy gaps, we present zoomed-in plots of the energy levels in Figure \ref{fig:energy_gaps_zoom} for a single realization of the samples. These plots focus on the energy levels within the range \([-4, 4]\) around the Fermi energy \(E_F = 0\), offering a more detailed visualization of the spectrum. Each panel in this figure corresponds to the same disorder strengths \(W\) as in Figure \ref{fig:energy_gaps} and highlights the fine structure of the energy levels. This enhanced resolution helps to better illustrate the spectrum characteristics and the distribution of energy levels in proximity to the Fermi level.

\begin{figure}
	\centering
	\begin{subfigure}{}%
		\includegraphics[width=0.23\textwidth]{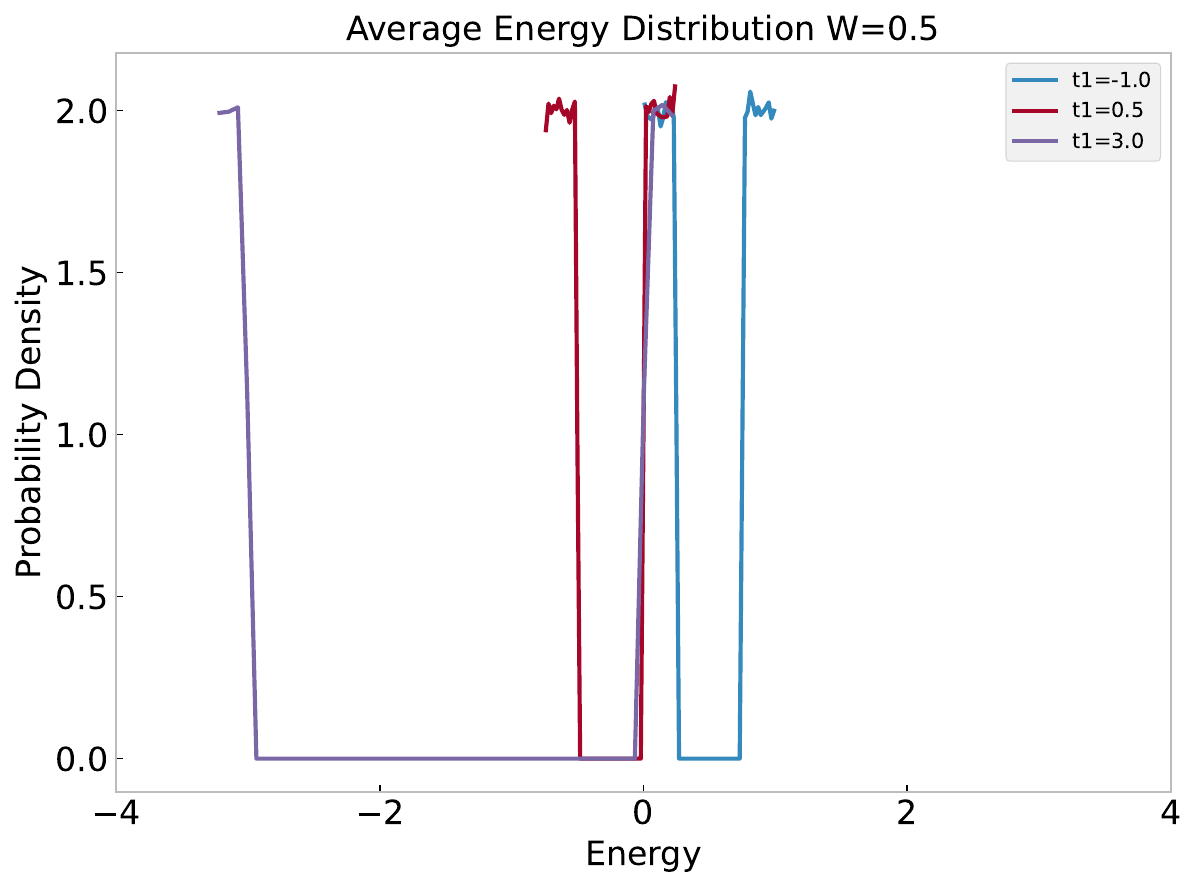}
	\end{subfigure}
	\begin{subfigure}{}%
		\includegraphics[width=0.23\textwidth]{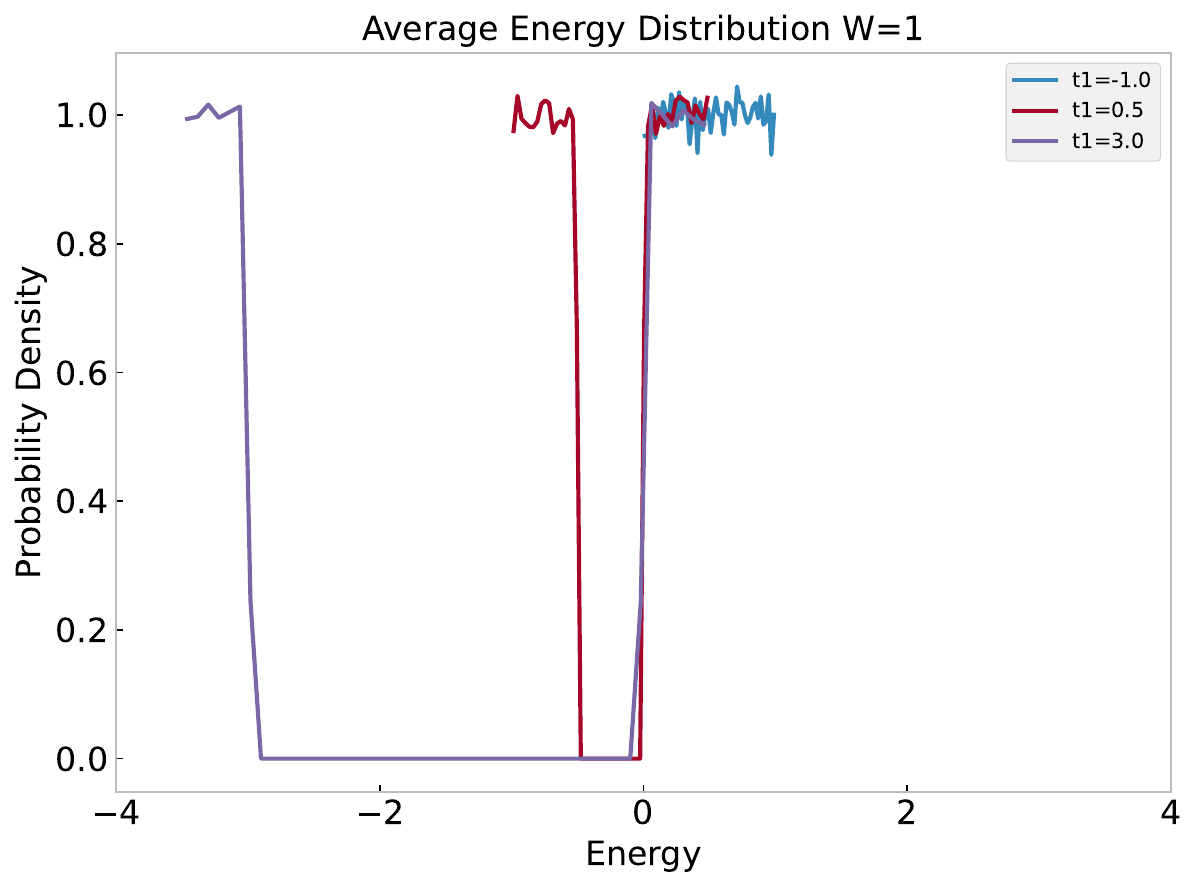}
	\end{subfigure}%
	\begin{subfigure}{}%
		\includegraphics[width=0.23\textwidth]{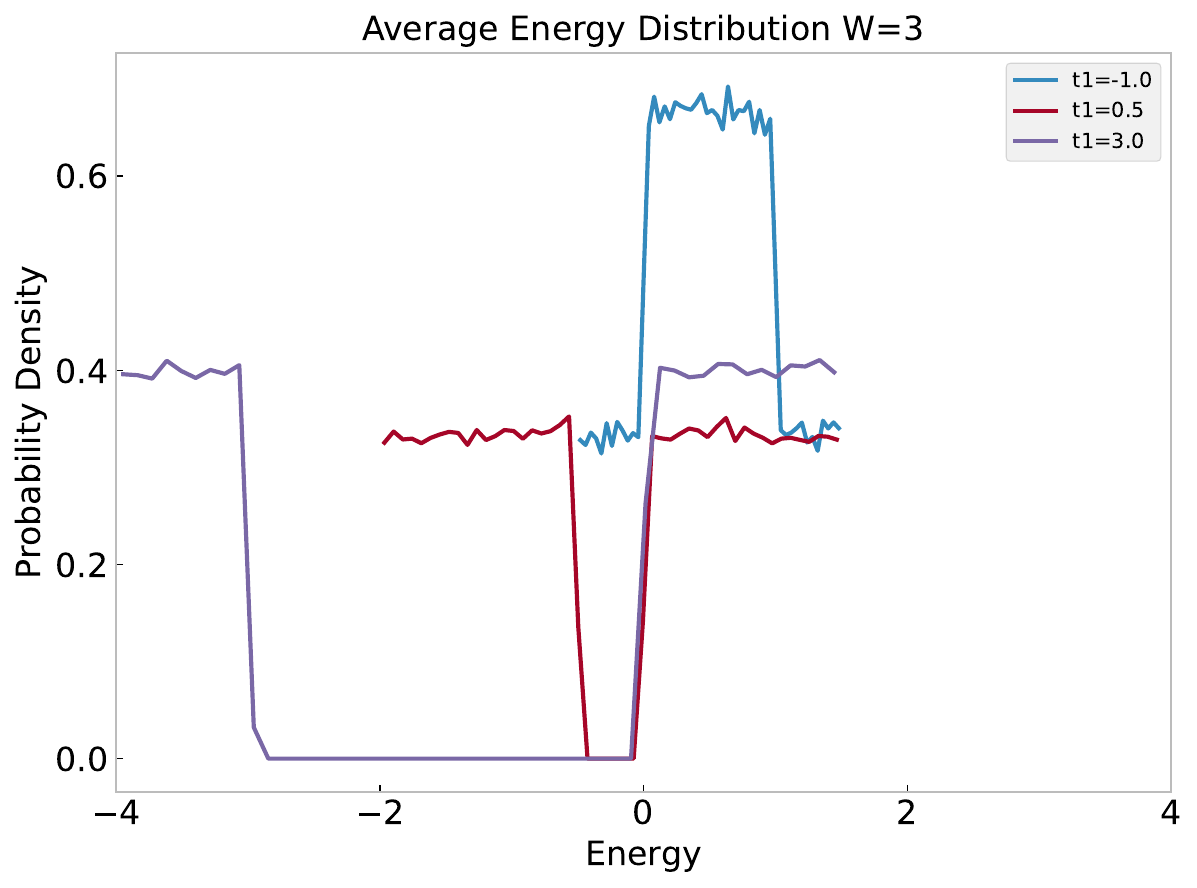}
	\end{subfigure}%
	\begin{subfigure}{}%
		\includegraphics[width=0.23\textwidth]{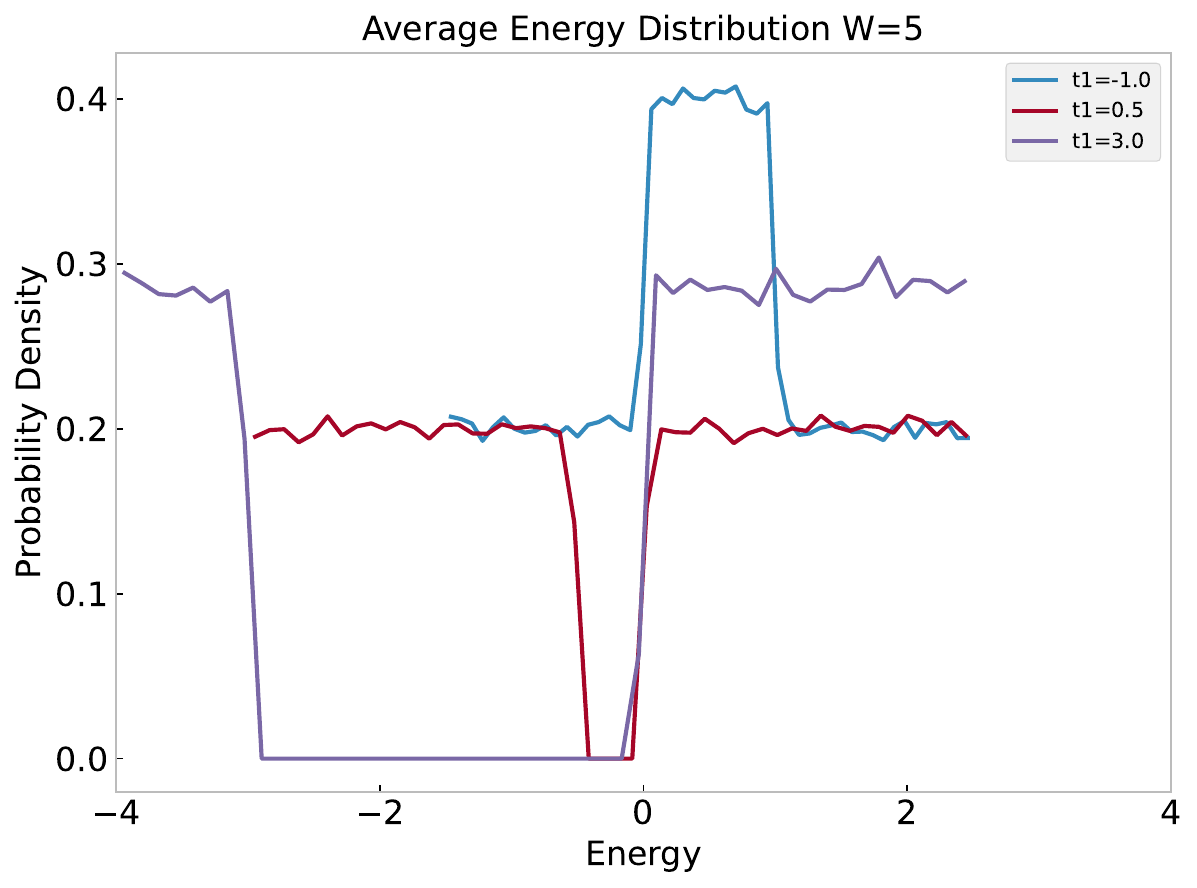}
	\end{subfigure}%
	\caption{Probability density distribution of the one-particle energies averaged over $200$ disorder realizations for system size $N=1000$ and various disorder strengths $W$. \label{fig:avg_energy_distribution}}
\end{figure}

To provide a more comprehensive understanding of the system's spectrum, we calculated the probability density of energy spectrum. For a system size of $N = 1000$ and disorder strengths $W$ as in Figure \ref{fig:energy_gaps}, we computed the probability density distribution of the one-particle energies, averaged over many realizations. The results, shown in Figure \ref{fig:avg_energy_distribution}, reveal a clear representation of the averaged energy distribution across different disorder strengths. The results are consistent with the previous figures.

\begin{figure}
	\centering
	\begin{subfigure}{}%
		\includegraphics[width=0.23\textwidth]{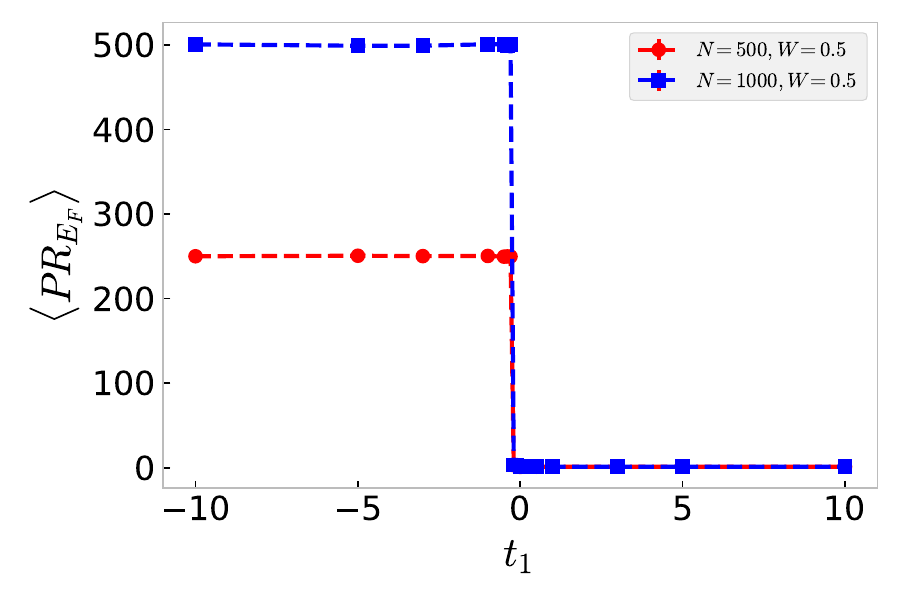}
	\end{subfigure}
	\begin{subfigure}{}%
		\includegraphics[width=0.23\textwidth]{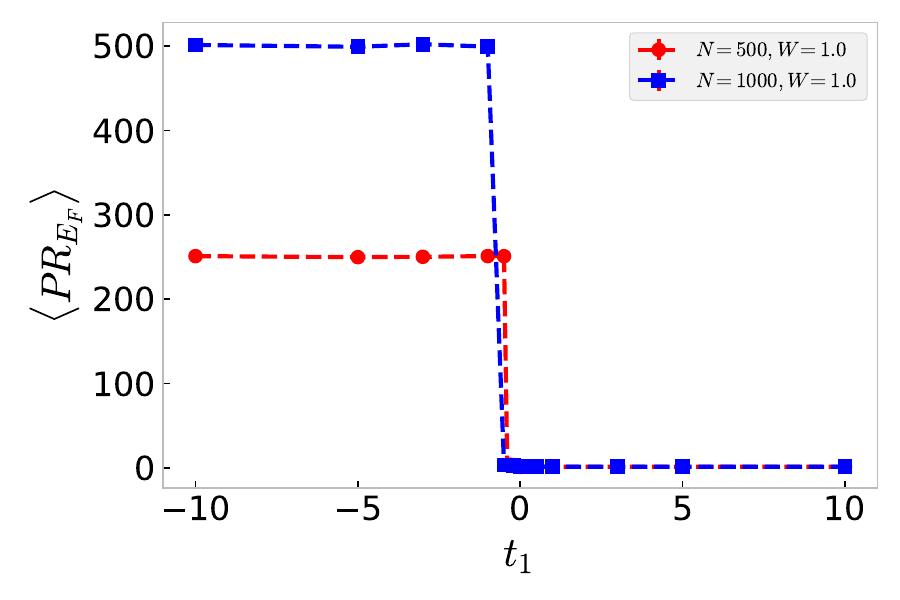}
	\end{subfigure}%
	\begin{subfigure}{}%
		\includegraphics[width=0.23\textwidth]{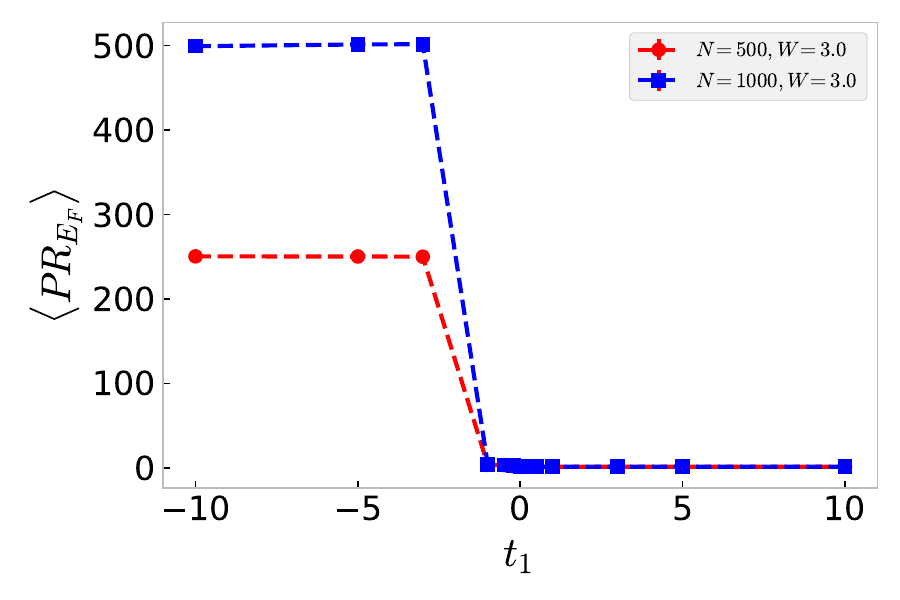}
	\end{subfigure}%
	\begin{subfigure}{}%
		\includegraphics[width=0.23\textwidth]{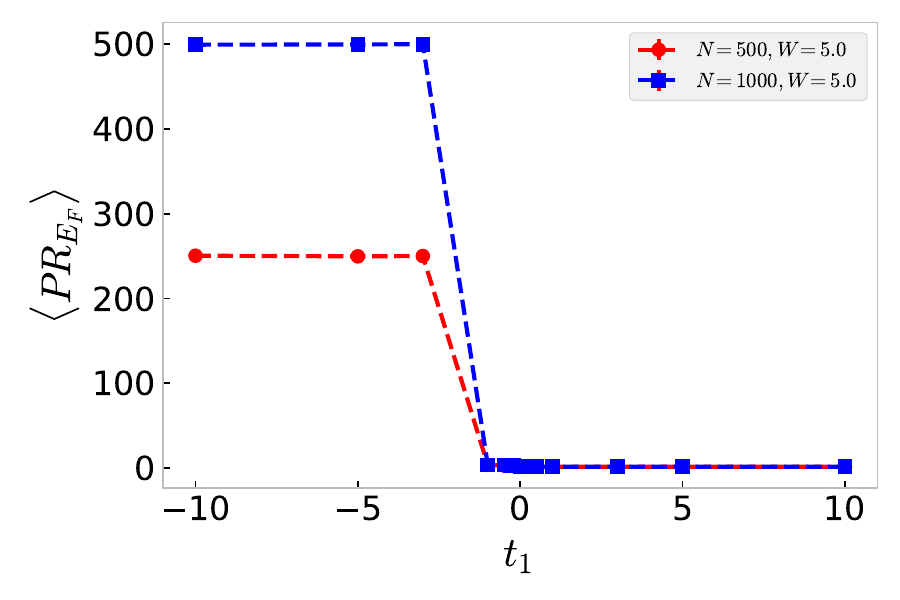}
	\end{subfigure}%
	\caption{Participation Ratio (PR) behavior in SLRTB model for varying hopping amplitude parameter $t$ and disorder strengths $W$. Each plot represents PR for system sizes $N = 500$ (markers (\CIRCLE)) and $N = 1000$ (markers ($\blacksquare$)), with data averaged over 200 samples. Error bars are significantly smaller than the PR values, indicating robust statistical results. PR values tend to be close to $N/2$ for negative hopping amplitudes ($t < 0$), except when the system is gapless, where PR equals 1. Conversely, PR consistently remains at 1 for positive hopping amplitudes ($t > 0$).\label{fig:participation_ratio} }
\end{figure}

In Figure \ref{fig:participation_ratio}, we illustrate the behavior of the Participation Ratio (PR) in the SLRTB model, characterized by different values of the hopping amplitude parameter \( t \) and disorder strengths \( W \). Each plot depicts PR for system sizes \( N = 500 \) and \( N = 1000 \), with the data averaged over 200 samples to ensure statistical reliability. Notably, the error bars in these plots are significantly smaller than the PR values themselves, indicating that the sample size is sufficient for robust statistical analysis. Our observations reveal distinct PR patterns: for negative hopping amplitudes (\( t < 0 \)), PR typically assumes a value close to \( N/2 \), where PR equals 1. In contrast, for positive hopping amplitudes (\( t > 0 \)), PR consistently remains at 1. However, PR deviates from this distinction in the region close to \( t = 0 \): as we increase \( W \), PR tends to approach 1 for larger magnitude negative \( t \) values, which contrasts with the gap behavior observed in Figure \ref{fig:energy_gaps}, where the gap exhibits similar behavior across different \( t \) values.

To analyze the scaling of the participation ratio (PR) with respect to critical exponents, we use the scaling relation:

\begin{equation}
\frac{PR}{N^{\alpha}} = f\left(N^{\beta} (t - t_c)\right),
\end{equation}

where $\alpha$ and $\beta$ are critical exponents, and $t_c$ is the critical value of the parameter \( t \) where the PR transitions from a finite value to 1. To find the critical exponents $\alpha$ and $\beta$, and the critical parameter \( t_c \), we use a cost function that quantifies the quality of the data collapse. To analyze the scaling of the participation ratio (PR), we first calculate the rescaled PR values for each system size \( N \) and parameter \( t \). This involves computing the PR and then rescaling it by dividing by \( N^{\alpha} \). Next, we define the scaling variable as \( N^{\beta} (t - t_c) \), which captures the influence of the system size and the critical parameter \( t_c \) on the PR.

The cost function \(\mathcal{C}_Q\) is then defined to measure the deviations in the scaled data. It is given by:

\begin{equation}
\mathcal{C}_Q = \frac{\sum_{i=1}^{N_p-1} |Q_{i+1} - Q_i|}{\max\{Q_i\} - \min\{Q_i\}} - 1,
\end{equation}

where \( Q_i \) represents the rescaled PR values, and \( N_p \) is the total number of data points. The values of \( Q_i \) are sorted according to the scaling variable \( N^{\beta} (t - t_c) \), arranged in non-decreasing order.

\begin{figure}
	\centering
	\begin{subfigure}{}%
		\includegraphics[width=0.23\textwidth]{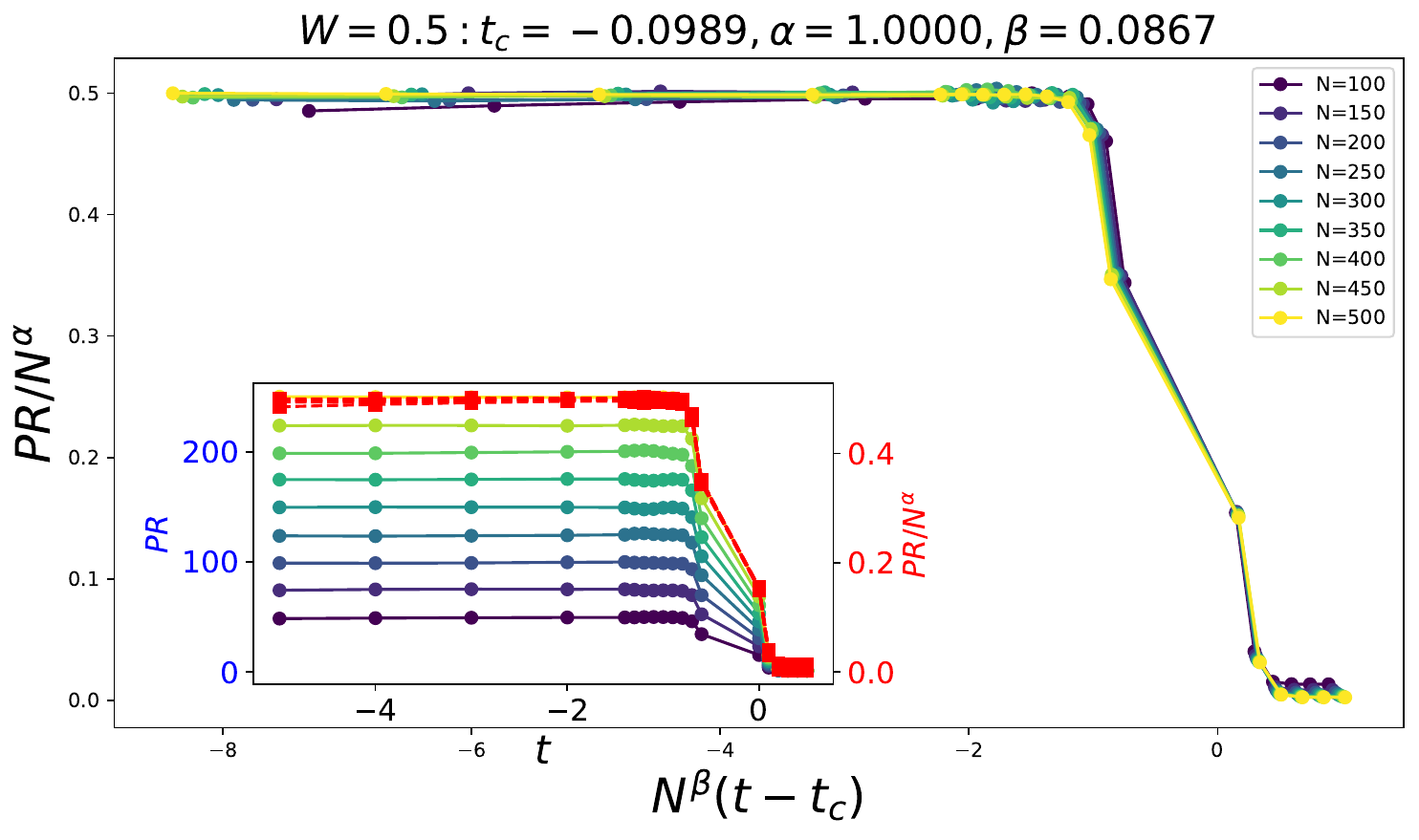}
	\end{subfigure}
	\begin{subfigure}{}%
		\includegraphics[width=0.23\textwidth]{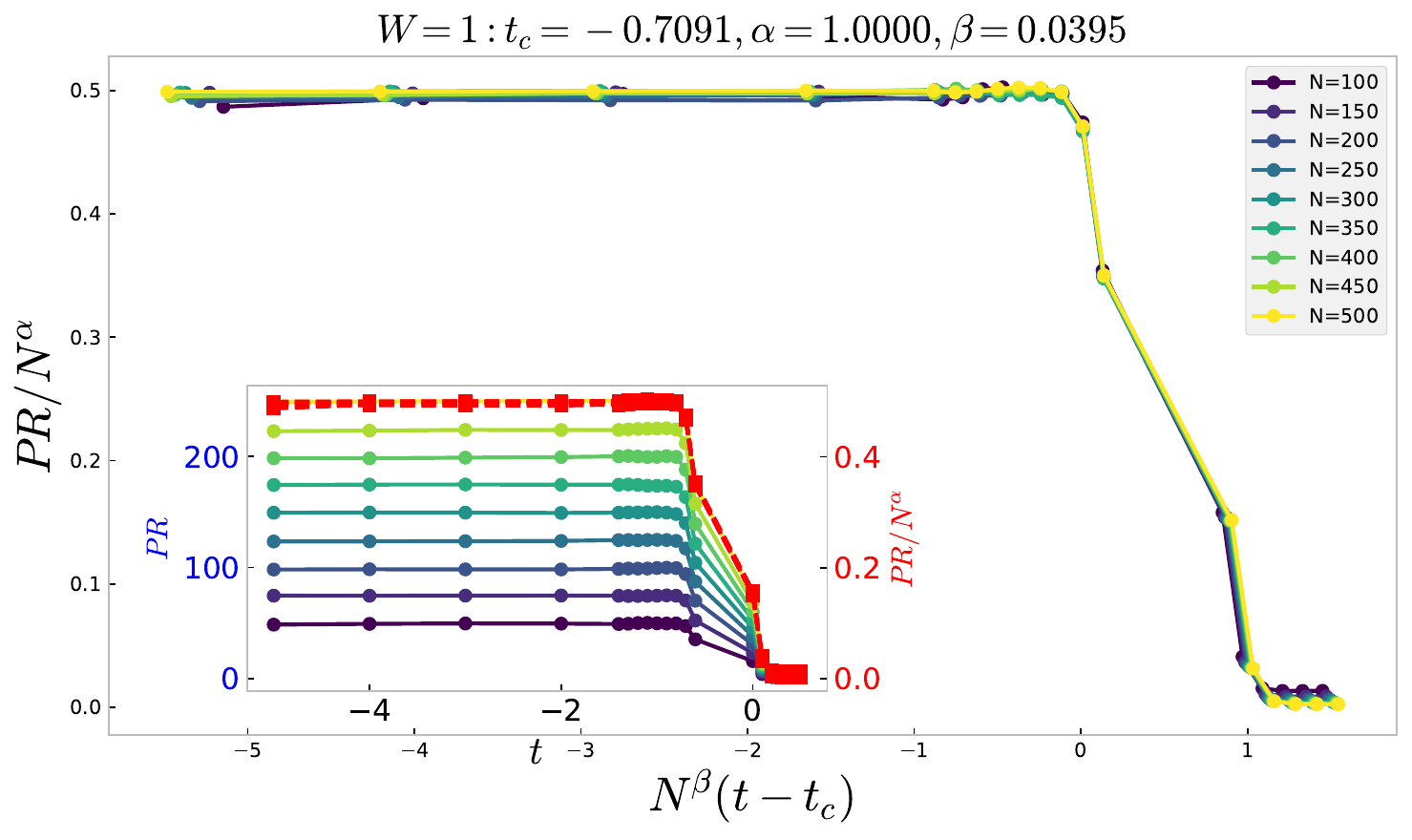}
	\end{subfigure}%
	\begin{subfigure}{}%
		\includegraphics[width=0.23\textwidth]{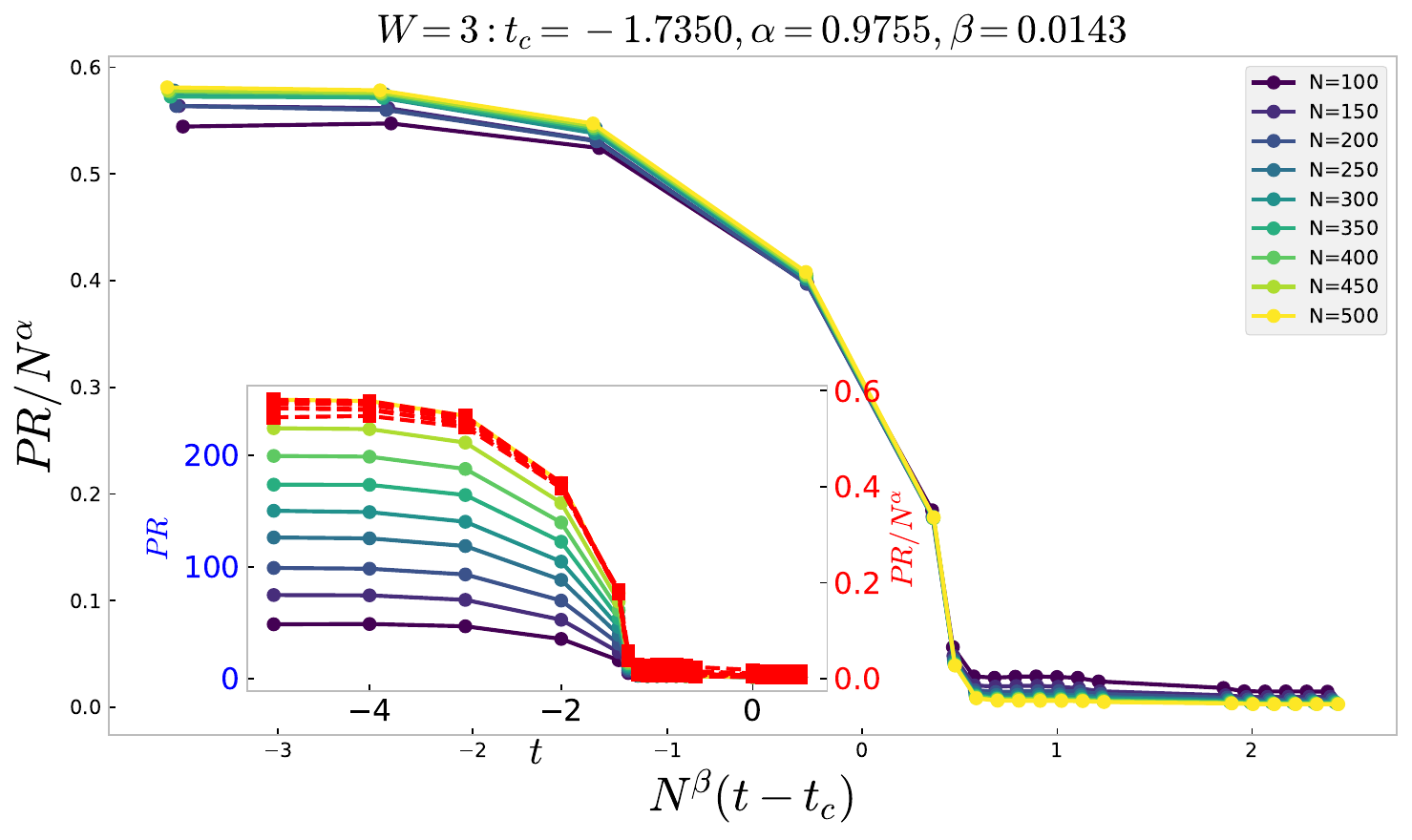}
	\end{subfigure}%
	\begin{subfigure}{}%
		\includegraphics[width=0.23\textwidth]{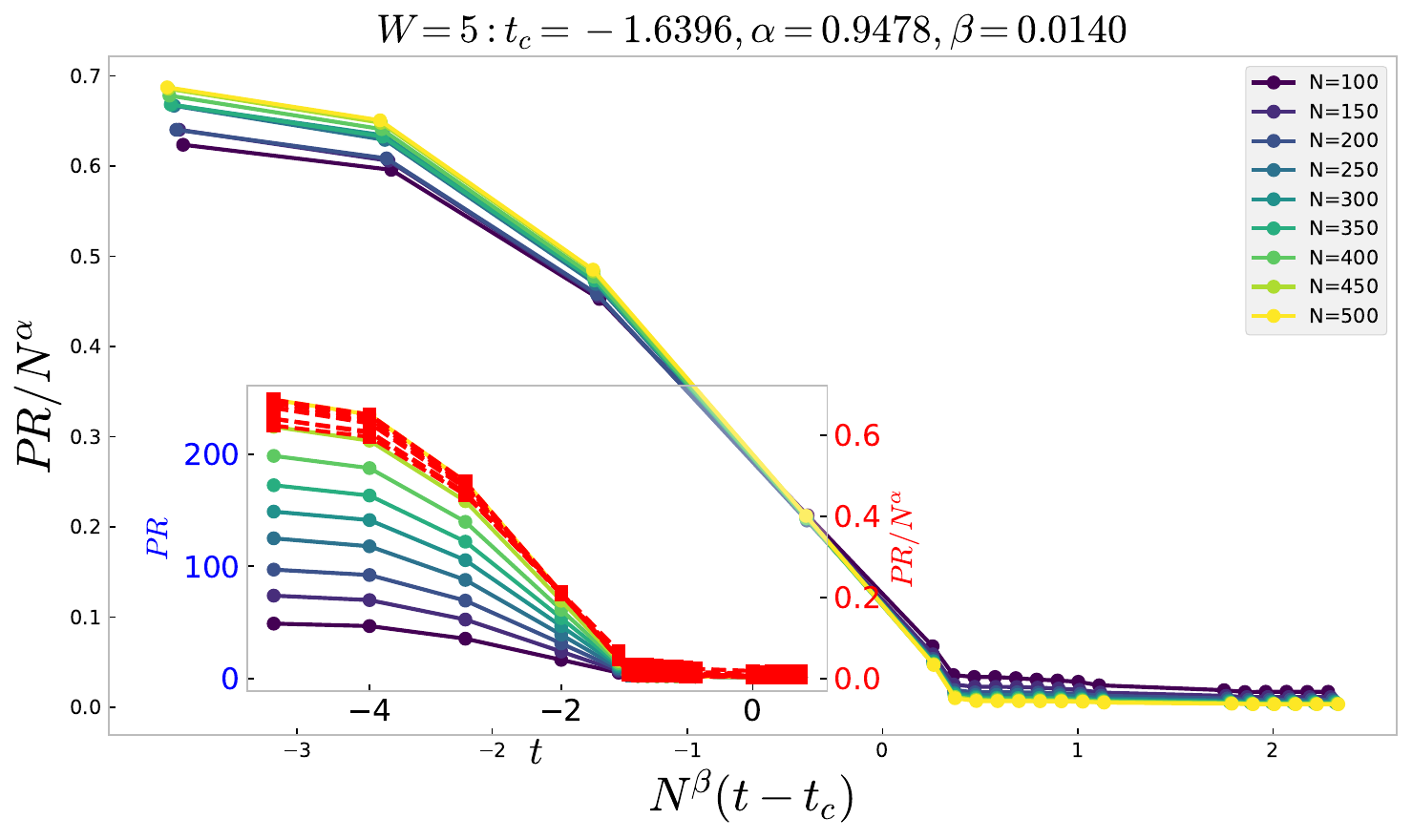}
	\end{subfigure}%
	\caption{The main plot shows the scaling collapse of the participation ratio (PR) for various system sizes \( N \) and disorder strengths \( W \). The rescaled PR values \( PR / N^\alpha \) are plotted as a function of the scaling variable \( N^\beta (t - t_c) \), with each curve corresponding to a different system size. Colors represent different \( N \) values, and the PR values are averaged over $200$ samples. The inset plot, located in the lower left, shows both the unscaled PR (left y-axis, blue) and the rescaled PR \( PR / N^\alpha \) (right y-axis, red) as a function of \( t \). The optimal critical parameters \( t_c \), \( \alpha \), and \( \beta \) for the disorder strength \( W \) are provided in the title.}
\label{fig:scalingPR}
\end{figure}

To determine the optimal values for the critical exponents \(\alpha\) and \(\beta\), as well as the critical parameter \( t_c \), we employ an optimization procedure. Initially, broad bounds for \(\alpha\), \(\beta\), and \( t_c \) are used to perform the optimization, providing rough estimates. Based on these initial estimates, the bounds are refined to focus on a smaller, more relevant region of the parameter space. Multiple optimization runs are then carried out to explore various regions and increase the likelihood of finding the global minimum. The optimal values of $\alpha$, $\beta$, and \( t_c \) are determined from the run with the lowest cost function value. The success of the data collapse is visualized by plotting the rescaled PR against the scaling variable \( N^{\beta} (t - t_c) \). A successful collapse indicates that data points from different system sizes align well when plotted against the scaling variable. These calculations have been performed for the dataset shown in Figure \ref{fig:energy_gaps}, and the results of the scaling analysis are presented in Figure \ref{fig:scalingPR}. As \( W \) increases, we observe that the transition point \( t_c \) shifts towards more negative values. This behavior is consistent with the visual evidence presented in Fig.~\ref{fig:participation_ratio}. The observed shift is attributed to the fact that as \( W \) increases, the system becomes gapless at increasingly larger negative values of \( t \).

\begin{figure}
	\centering
	\begin{subfigure}{}%
		\includegraphics[width=0.23\textwidth]{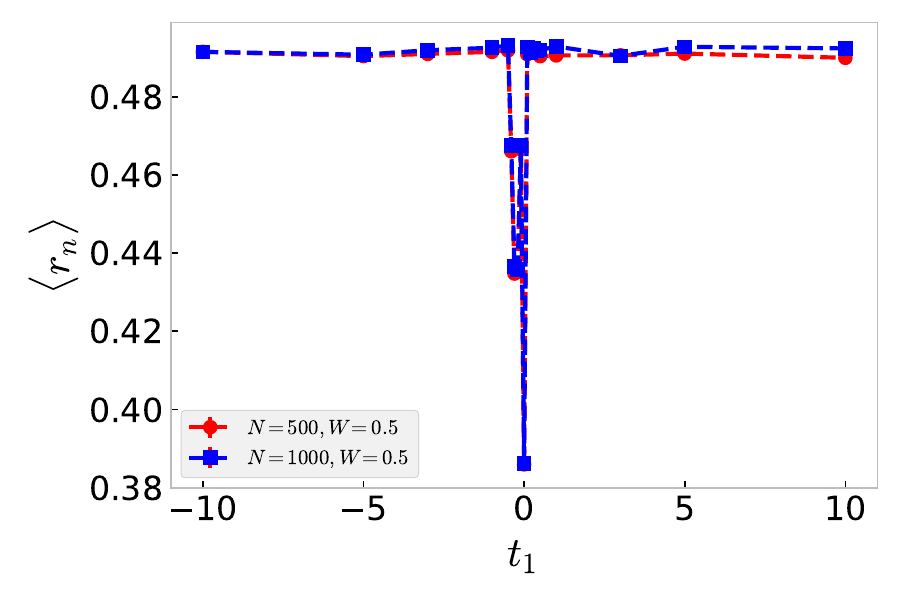}
	\end{subfigure}
	\begin{subfigure}{}%
		\includegraphics[width=0.23\textwidth]{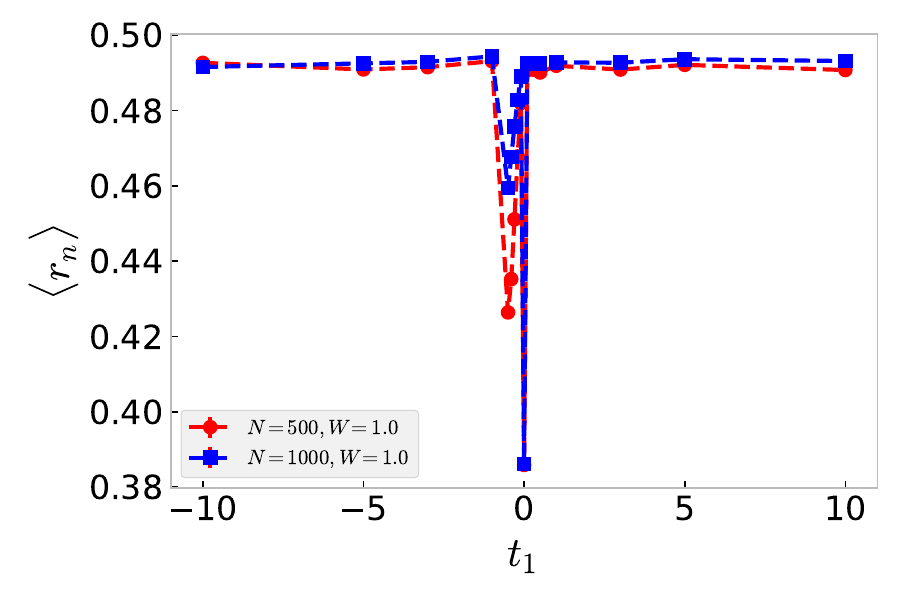}
	\end{subfigure}%
	\begin{subfigure}{}%
		\includegraphics[width=0.23\textwidth]{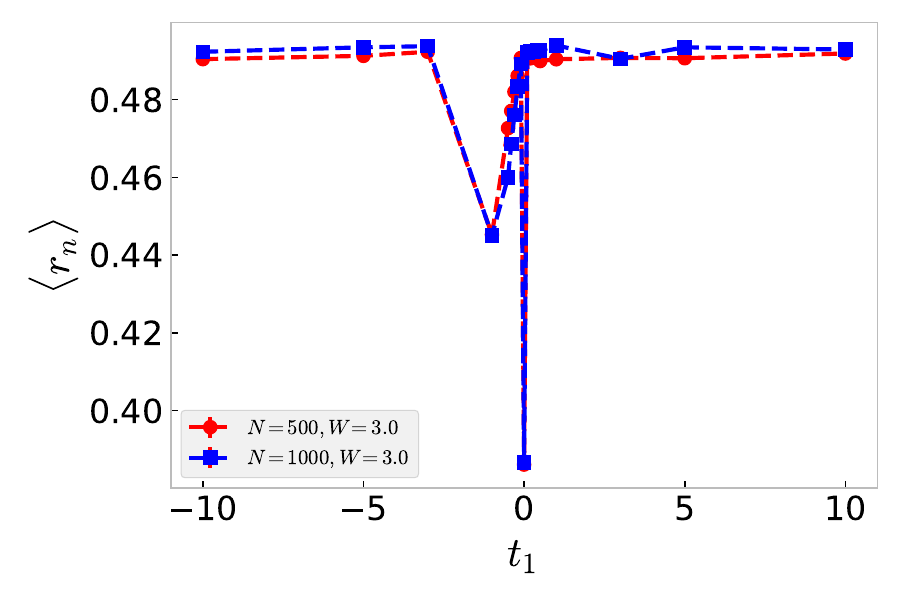}
	\end{subfigure}%
	\begin{subfigure}{}%
		\includegraphics[width=0.23\textwidth]{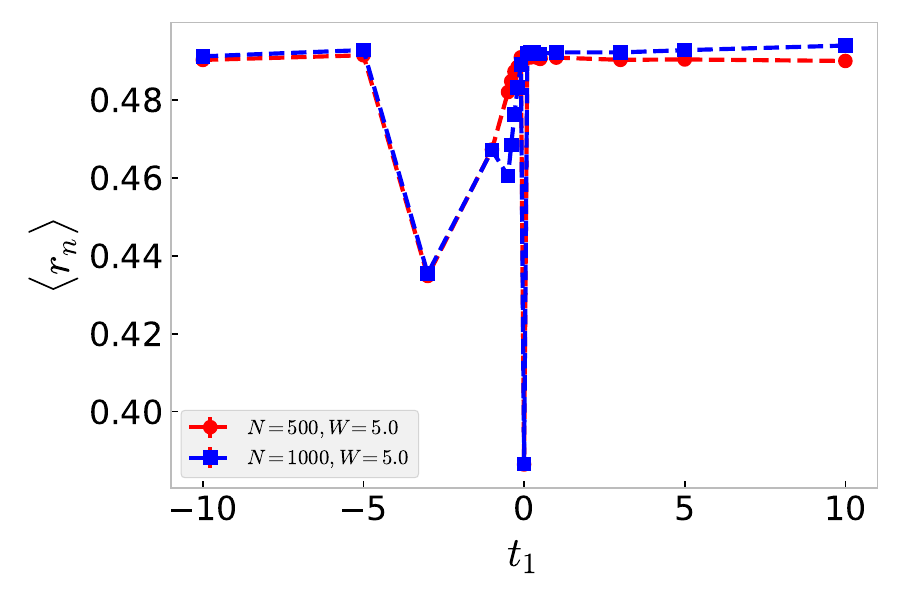}
	\end{subfigure}%
	\caption{Behavior of the ratio of adjacent level spacings ($r_n$) in SLRTB model for varying hopping amplitude parameter $t$. Each plot represents $r_n$ for system sizes $N = 500$ (markers (\CIRCLE)) and $N = 1000$ (markers ($\blacksquare$)), with data averaged over 500 samples. Error bars, representing sample averaging over 500 samples, are remarkably small, underscoring the consistency of these results.\label{fig:rn} }
\end{figure}

The behavior of the ratio of level spacings ($r_n$) is a crucial indicator of the transition in SLRTB model. For $t_{ij} = 0$, corresponding to the condition of zero hopping amplitudes, $r_n$ is approximately 0.39. In contrast, when $t_{ij}$ is not equal to zero ($t_{ij} \neq 0$), $r_n$ increases to around 0.49. Remarkably, this behavior remains consistent for different system sizes, N=500 and N=1000, as illustrated in the plots (see Fig. \ref{fig:rn})

\begin{figure}
	\centering
	\begin{subfigure}{}%
		\includegraphics[width=0.23\textwidth]{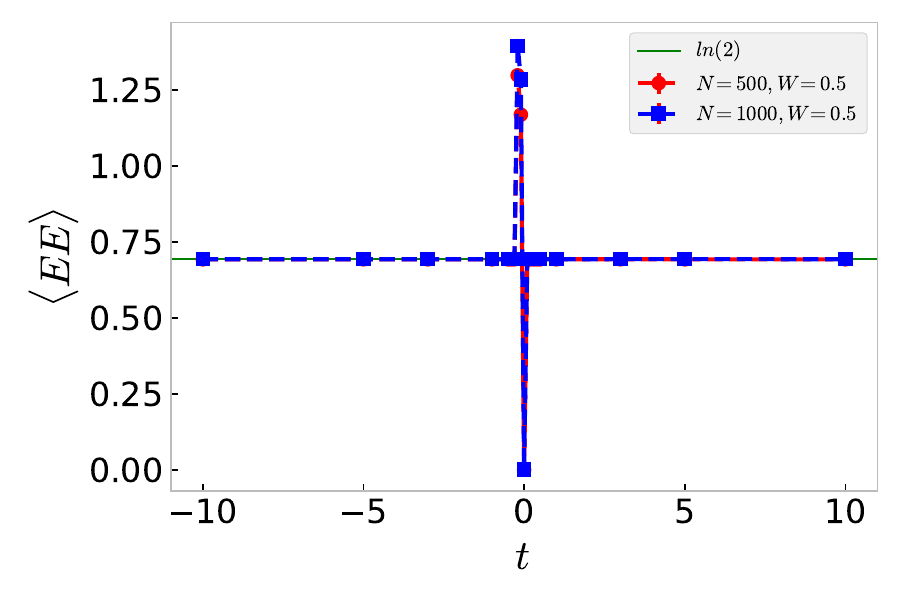}
	\end{subfigure}
	\begin{subfigure}{}%
		\includegraphics[width=0.23\textwidth]{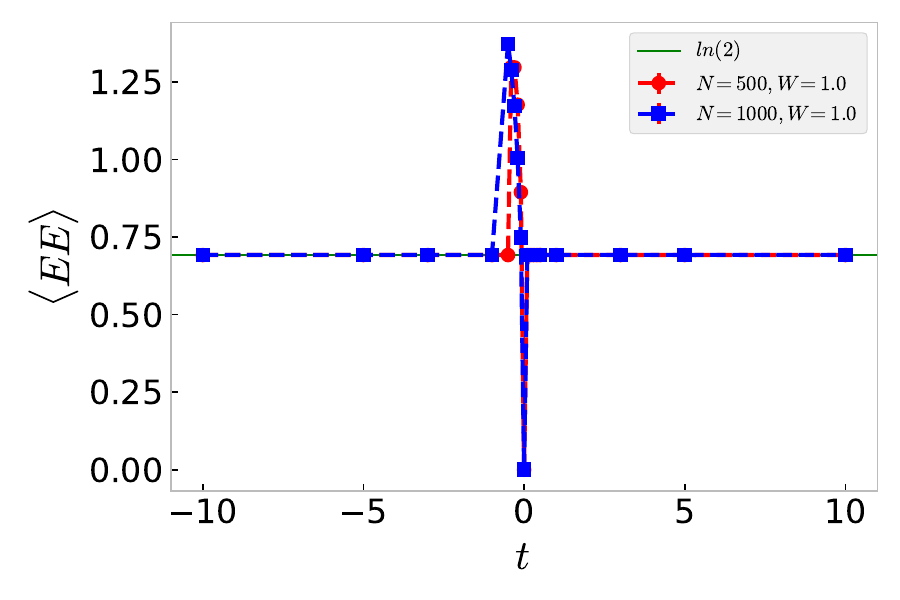}
	\end{subfigure}%
	\begin{subfigure}{}%
		\includegraphics[width=0.23\textwidth]{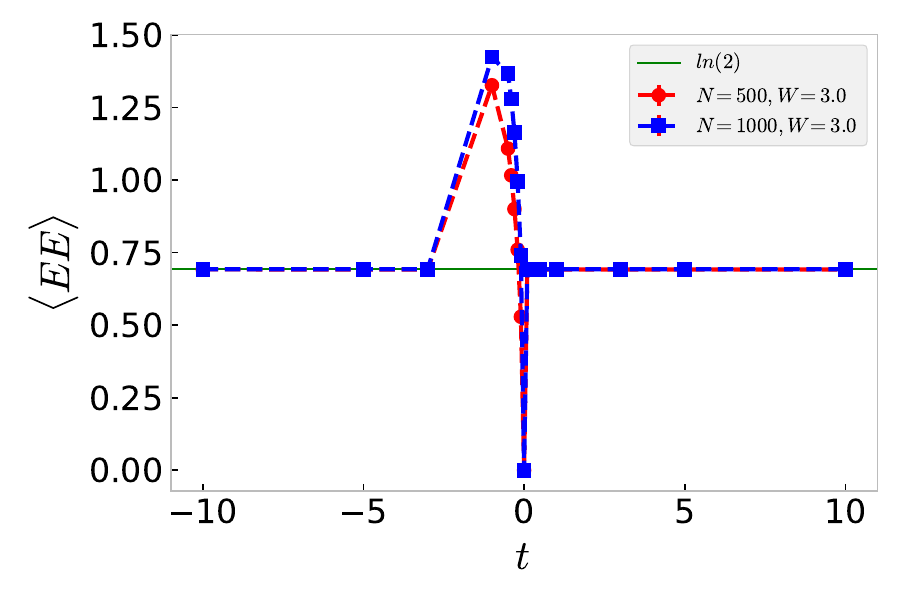}
	\end{subfigure}%
	\begin{subfigure}{}%
		\includegraphics[width=0.23\textwidth]{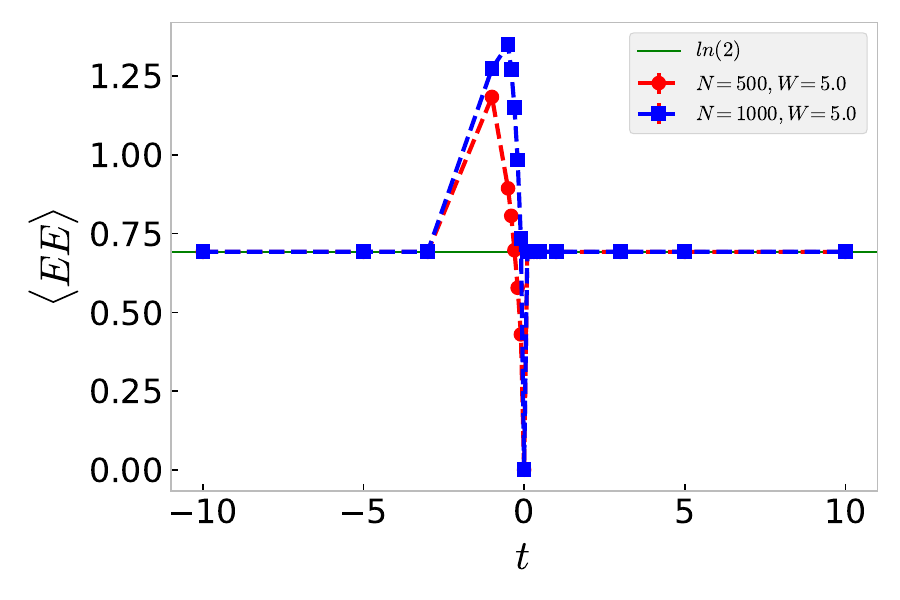}
	\end{subfigure}%
	\caption{Entanglement Entropy (EE) for the Selective Long-Range Hamiltonian. The figure illustrates the behavior of EE as a function of the hopping amplitude parameter $t$ for varying disorder strengths $W$. Each subplot represents a different disorder strength: (a) $W = 0.5$, (b) $W = 1$, (c) $W = 3$, and (d) $W = 5$. Results are shown for two system sizes, $N = 500$ (markers (\CIRCLE)) and $N = 1000$  (markers ($\blacksquare$)), and each data point is obtained through disorder averaging over 500 samples. Error bars, though often minuscule, are included to emphasize the reliability of the results. The horizontal line at $EE \approx ln(2) \approx 0.69$ serves as a reference value.\label{fig:EE} }
\end{figure}

In the following, we delve into the analysis of entanglement entropy (EE) in our model, shedding light on the intricate interplay between electron entanglement and localization. Our numerical results for EE, which are presented in Figure \ref{fig:EE}. In SLRTB model, we have observed intriguing behavior in the entanglement entropy (EE) as a function of the hopping amplitude parameter $t$. For most of the parameter space, EE converges to a value of approximately ln(2), which is approximately 0.69 corresponding to one bit of information. This value signifies a characteristic level of entanglement in the system. However, there are exceptions to this pattern. In particular, in regions where $t_{ij} < 0$ and there is an increase in the number of filled bands ($N_F$), the EE exceeds the $\ln(2)$ threshold, indicating enhanced entanglement. Notably, when $t_{ij} = 0$, corresponding to the case of zero hopping amplitudes, EE drops to zero, signifying minimal entanglement.

These observations highlight the role of energy gaps and the sign of hopping amplitudes in influencing the entanglement properties of the system. While most of the parameter space maintains a relatively constant level of entanglement, gappless regions with $t_{ij} < 0$ exhibit heightened entanglement.

To investigate the finite-size scaling of the entanglement entropy (EE) for small negative values of the hopping amplitude \( t \), we applied the cost function method previously used for the participation ratio (PR) analysis. In this approach, we optimized the critical parameters, including the critical hopping amplitude \( t_c \) and the critical exponent \( \nu \), to achieve the best scaling collapse of the EE data across different system sizes. The cost function was minimized to ensure the optimal alignment of the rescaled EE curves, allowing us to extract meaningful information about the critical behavior in the vicinity of the observed peak in the EE. The results of this scaling analysis demonstrate a clear critical behavior in the system for small negative values of \( t \), confirming the sensitivity of the EE to system size in this regime. As the disorder strength \( W \) increases, we observe that the critical point \( t_c \) shifts to more negative values in the scaling of the entanglement entropy. This trend aligns with the behavior observed in Fig.~\ref{fig:EE}. The shift can be explained by the system becoming increasingly gapless at larger negative values of \( t \) as \( W \) grows, which affects the entanglement properties accordingly.

\begin{figure}
	\centering
	\begin{subfigure}{}%
		\includegraphics[width=0.23\textwidth]{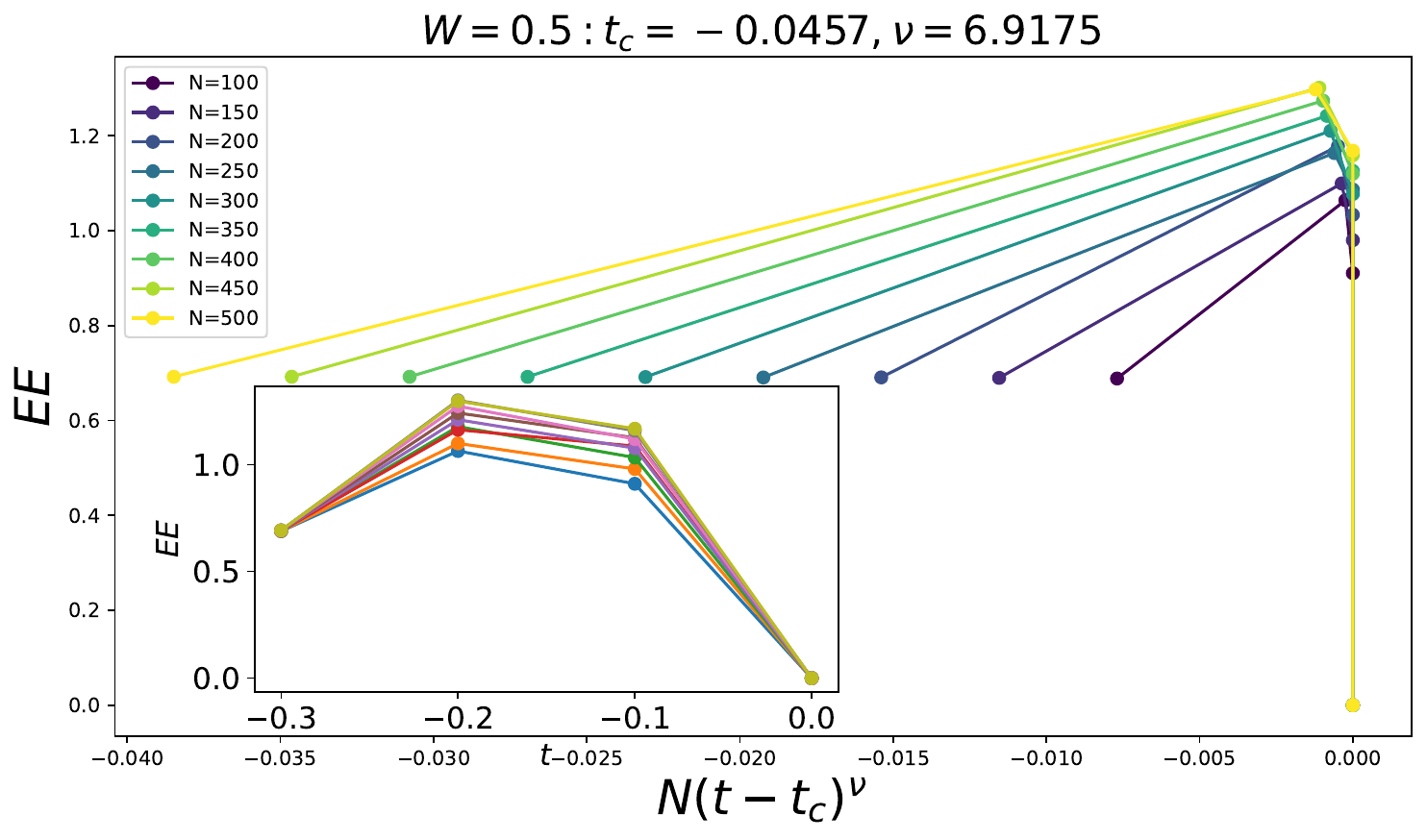}
	\end{subfigure}
	\begin{subfigure}{}%
		\includegraphics[width=0.23\textwidth]{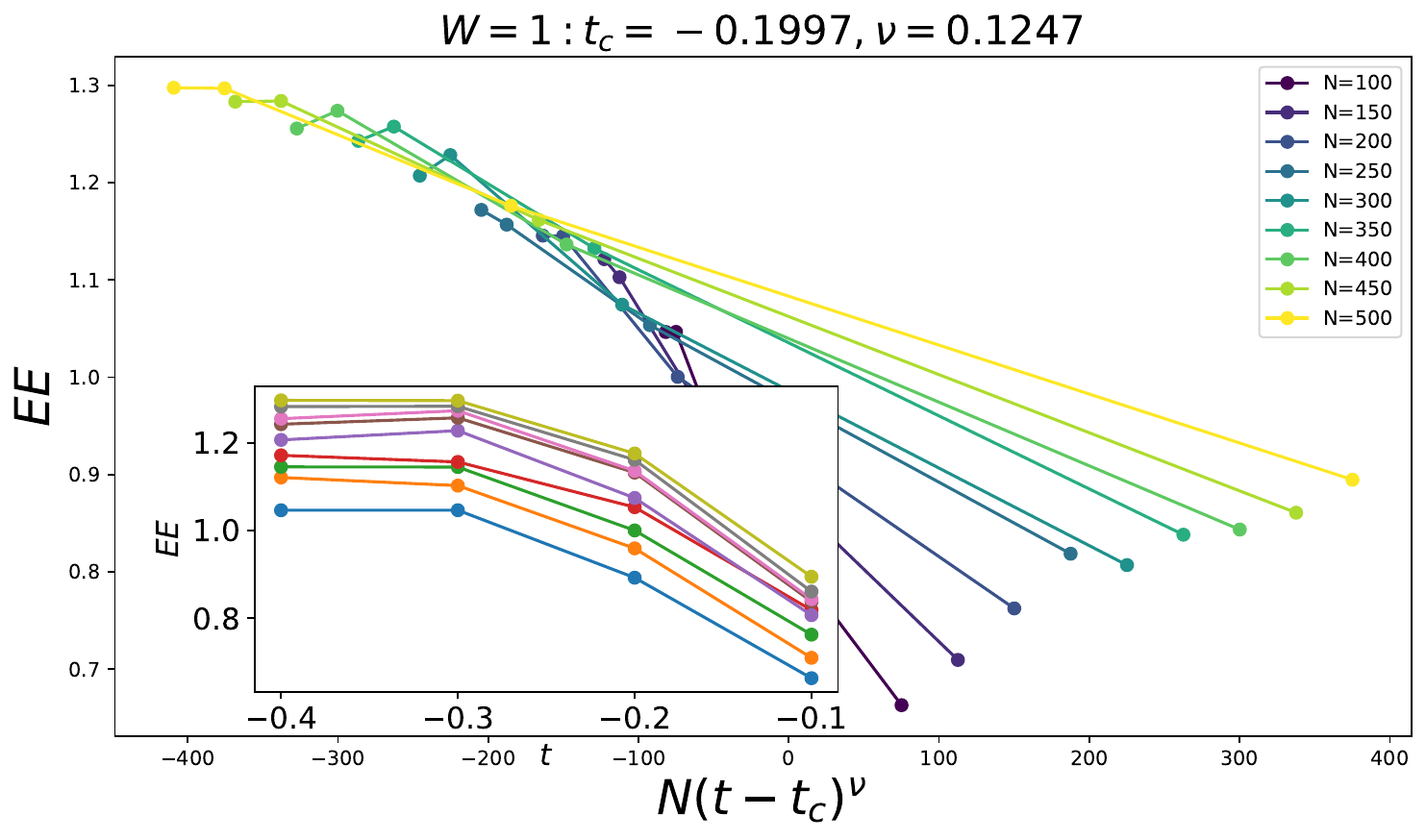}
	\end{subfigure}%
	\begin{subfigure}{}%
		\includegraphics[width=0.23\textwidth]{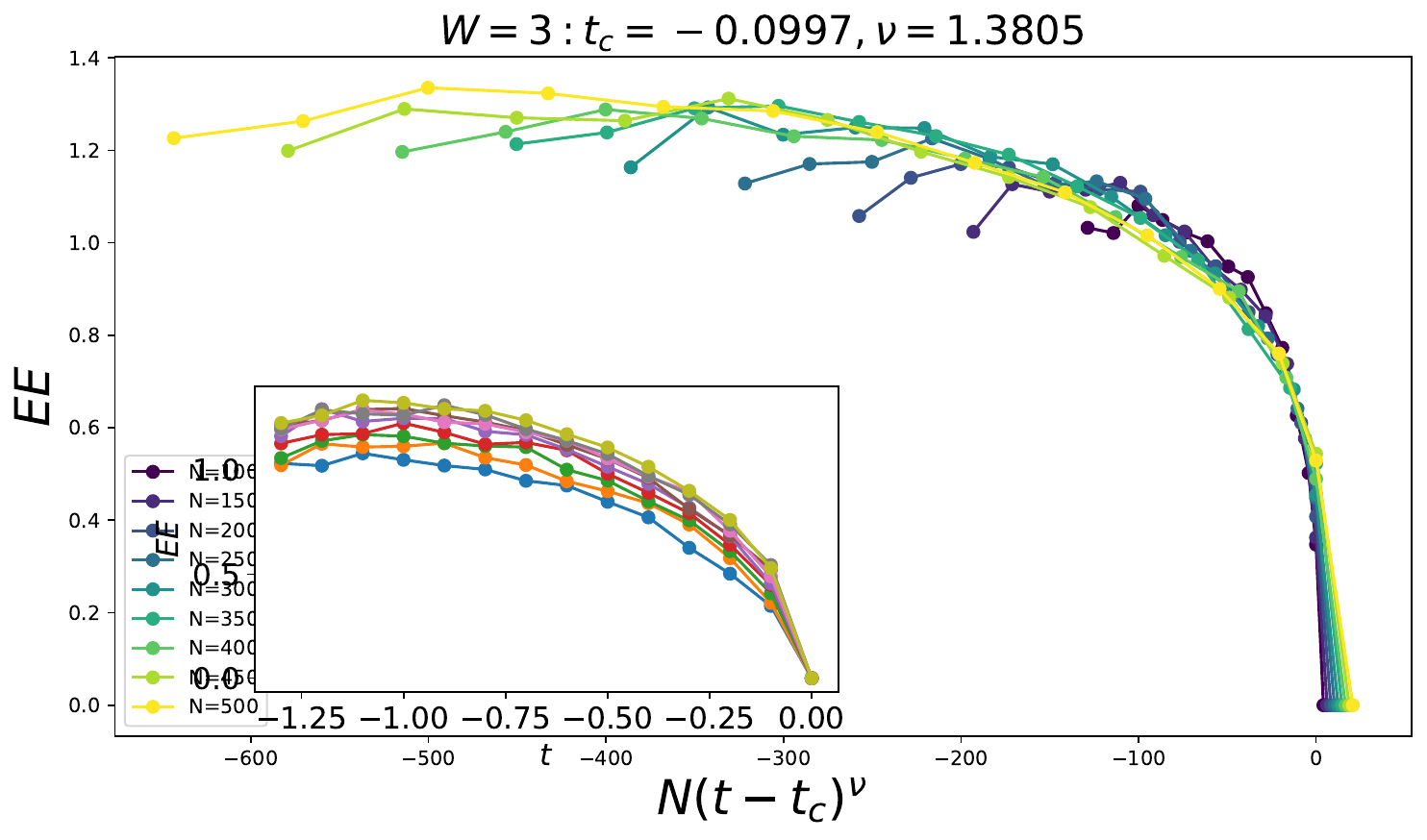}
	\end{subfigure}%
	\begin{subfigure}{}%
		\includegraphics[width=0.23\textwidth]{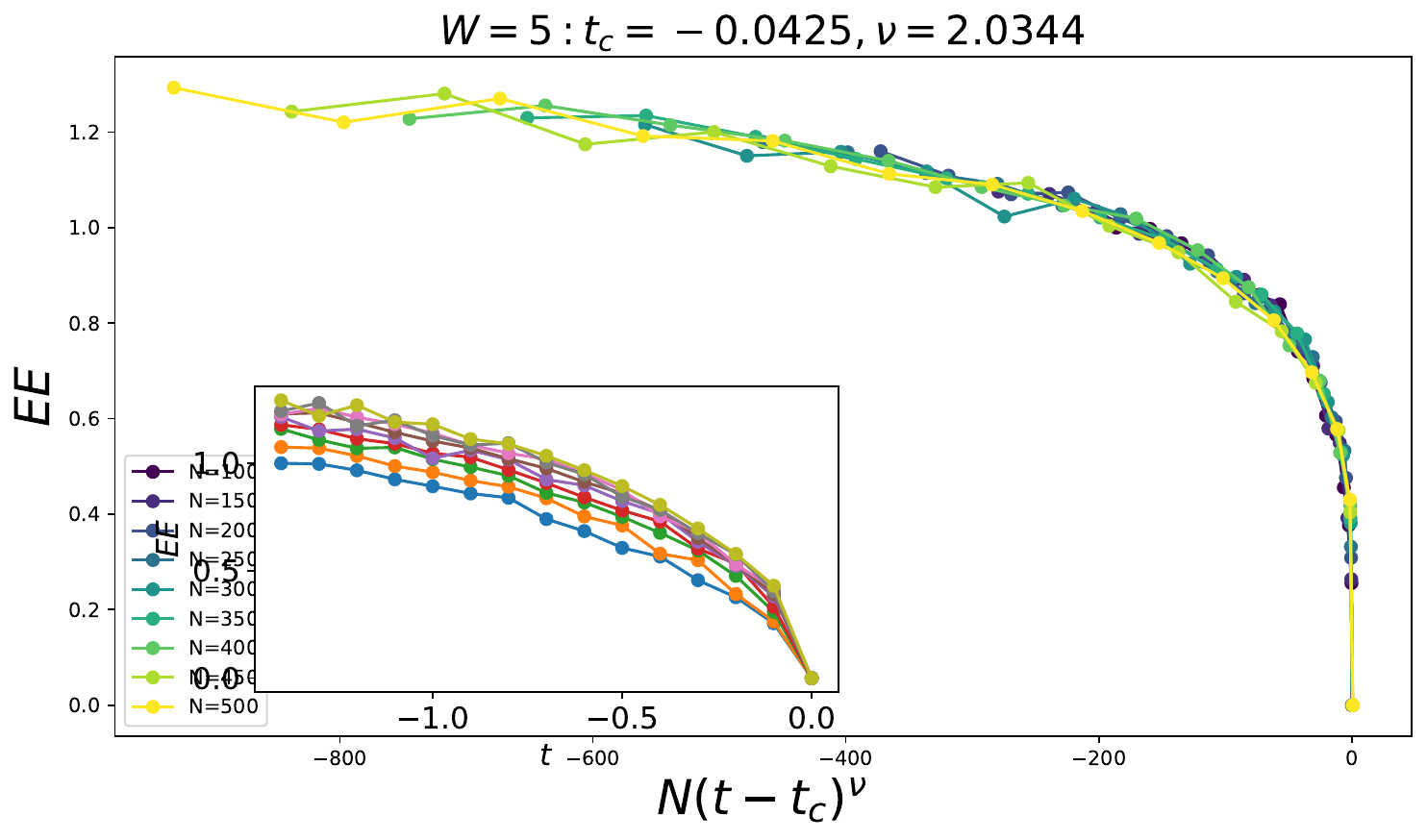}
	\end{subfigure}%
	\caption{Scaling collapse of the entanglement entropy (EE) for various system sizes \( N \) and disorder strengths \( W \). The main plot shows the EE values as a function of the scaling variable \( N (t - t_c)^\nu \). Each curve corresponds to a different system size, with the colors indicating different \( N \) values. The inset plot shows the unscaled EE as a function of the hopping parameter \( t \). The critical parameters \( t_c \), and \(\nu \), are annotated in the title. The EE values are averaged over $200$ disorder realizations. \label{fig:scalingEE}
}
\end{figure}

\section{Conclusion}\label{conclusion}

In this study, we have thoroughly explored the complex interplay between disorder-induced electron localization and long-range hopping amplitudes within the framework of the Selective Long-Range Tight-Binding Model (SLRTB). Our investigation has yielded several key insights into the behavior of electrons in such a disordered system.

We began by analyzing the energy spectrum, particularly focusing on energy gaps at the Fermi level \( E_F = 0 \). These gaps served as indicators of localized and delocalized phases in the system. Our results demonstrated a clear transition from gapless to gapped phases as the hopping amplitude \( t \) and disorder strength \( W \) varied. This transition highlighted the system's sensitivity to these parameters, revealing intriguing patterns in energy level distribution.

Next, we examined the Participation Ratio (PR) to further probe localization behavior. Our numerical simulations indicated that for negative hopping amplitudes (\( t < 0 \)), the PR values typically approach \( N/2 \), except in regions where the system exhibits gapless behavior, where the PR drops to 1. In contrast, for positive hopping amplitudes (\( t > 0 \)), the PR remains consistently at 1, signifying a lack of localization.

The analysis of the ratio of level spacings (\( r_n \)) provided additional insights into the transition within the SLRTB model. For \( t_{ij} = 0 \), \( r_n \) was approximately 0.39, characteristic of localized states. As \( t_{ij} \) deviated from zero, \( r_n \) increased to around 0.49, indicating a transition to delocalized behavior. Importantly, this behavior was consistent across different system sizes, reinforcing the robustness of our results.

We also performed a finite-size scaling analysis on the entanglement entropy (EE), focusing on the region of small negative \( t \), where the EE exhibited significant variation. The scaling analysis revealed that as \( W \) increases, the critical point \( t_c \) shifts to more negative values, mirroring the behavior observed in the PR.

An alternative to our model could be a model where all sites are connected, regardless of the sign of the on-site energies, or a model with only quarter of the sites connected. However, these models exhibit completely trivial behavior and do not capture the distinctive features observed in our proposed model.

In conclusion, the SLRTB model reveals a rich landscape of electron behavior under the influence of disorder and long-range hopping. The observed transitions between gapped and gapless phases, and the associated changes in localization, provide valuable insights into the fundamental properties of disordered systems. Additionally, incorporating electron-electron interactions into this model could pave the way for studying many-body localization \cite{alet2018many, pal2010many, abanin2019colloquium, iyer2013many}. Moreover, a comparative analysis with other disordered models, such as the Aubry-André model \cite{biddle2009localization, an2021interactions, mastropietro2015localization, ray2018drive, bu2022quantum}, could further elucidate the unique features of the SLRTB model and its broader applicability in quantum materials.

\section*{Data Availability Statement} 

The data that support the findings of this study are available from the corresponding author, upon reasonable request.

\acknowledgments

The author gratefully acknowledge the high performance
computing center of university of Mazandaran for providing computing
resource and time. This research was financed by a research grant from the University of Mazandaran

\bibliographystyle{apsrev4-2.bst}
\bibliography{/home/cmp/Dropbox/physics/Bib/reference.bib}

\begin{thebibliography}{73}%
\makeatletter
\providecommand \@ifxundefined [1]{%
 \@ifx{#1\undefined}
}%
\providecommand \@ifnum [1]{%
 \ifnum #1\expandafter \@firstoftwo
 \else \expandafter \@secondoftwo
 \fi
}%
\providecommand \@ifx [1]{%
 \ifx #1\expandafter \@firstoftwo
 \else \expandafter \@secondoftwo
 \fi
}%
\providecommand \natexlab [1]{#1}%
\providecommand \enquote  [1]{``#1''}%
\providecommand \bibnamefont  [1]{#1}%
\providecommand \bibfnamefont [1]{#1}%
\providecommand \citenamefont [1]{#1}%
\providecommand \href@noop [0]{\@secondoftwo}%
\providecommand \href [0]{\begingroup \@sanitize@url \@href}%
\providecommand \@href[1]{\@@startlink{#1}\@@href}%
\providecommand \@@href[1]{\endgroup#1\@@endlink}%
\providecommand \@sanitize@url [0]{\catcode `\\12\catcode `\$12\catcode
  `\&12\catcode `\#12\catcode `\^12\catcode `\_12\catcode `\%12\relax}%
\providecommand \@@startlink[1]{}%
\providecommand \@@endlink[0]{}%
\providecommand \url  [0]{\begingroup\@sanitize@url \@url }%
\providecommand \@url [1]{\endgroup\@href {#1}{\urlprefix }}%
\providecommand \urlprefix  [0]{URL }%
\providecommand \Eprint [0]{\href }%
\providecommand \doibase [0]{https://doi.org/}%
\providecommand \selectlanguage [0]{\@gobble}%
\providecommand \bibinfo  [0]{\@secondoftwo}%
\providecommand \bibfield  [0]{\@secondoftwo}%
\providecommand \translation [1]{[#1]}%
\providecommand \BibitemOpen [0]{}%
\providecommand \bibitemStop [0]{}%
\providecommand \bibitemNoStop [0]{.\EOS\space}%
\providecommand \EOS [0]{\spacefactor3000\relax}%
\providecommand \BibitemShut  [1]{\csname bibitem#1\endcsname}%
\let\auto@bib@innerbib\@empty
\bibitem [{\citenamefont {Anderson}(1958)}]{PhysRev.109.1492}%
  \BibitemOpen
  \bibfield  {author} {\bibinfo {author} {\bibfnamefont {P.~W.}\ \bibnamefont
  {Anderson}},\ }\href {https://doi.org/10.1103/PhysRev.109.1492} {\bibfield
  {journal} {\bibinfo  {journal} {Phys. Rev.}\ }\textbf {\bibinfo {volume}
  {109}},\ \bibinfo {pages} {1492} (\bibinfo {year} {1958})}\BibitemShut
  {NoStop}%
\bibitem [{\citenamefont {Lagendijk}\ \emph {et~al.}(2009)\citenamefont
  {Lagendijk}, \citenamefont {Van~Tiggelen},\ and\ \citenamefont
  {Wiersma}}]{lagendijk2009fifty}%
  \BibitemOpen
  \bibfield  {author} {\bibinfo {author} {\bibfnamefont {A.}~\bibnamefont
  {Lagendijk}}, \bibinfo {author} {\bibfnamefont {B.}~\bibnamefont
  {Van~Tiggelen}},\ and\ \bibinfo {author} {\bibfnamefont {D.~S.}\ \bibnamefont
  {Wiersma}},\ }\href@noop {} {\bibfield  {journal} {\bibinfo  {journal} {Phys.
  Today}\ }\textbf {\bibinfo {volume} {62}},\ \bibinfo {pages} {24} (\bibinfo
  {year} {2009})}\BibitemShut {NoStop}%
\bibitem [{\citenamefont {Semeghini}\ \emph {et~al.}(2015)\citenamefont
  {Semeghini}, \citenamefont {Landini}, \citenamefont {Castilho}, \citenamefont
  {Roy}, \citenamefont {Spagnolli}, \citenamefont {Trenkwalder}, \citenamefont
  {Fattori}, \citenamefont {Inguscio},\ and\ \citenamefont
  {Modugno}}]{semeghini2015measurement}%
  \BibitemOpen
  \bibfield  {author} {\bibinfo {author} {\bibfnamefont {G.}~\bibnamefont
  {Semeghini}}, \bibinfo {author} {\bibfnamefont {M.}~\bibnamefont {Landini}},
  \bibinfo {author} {\bibfnamefont {P.}~\bibnamefont {Castilho}}, \bibinfo
  {author} {\bibfnamefont {S.}~\bibnamefont {Roy}}, \bibinfo {author}
  {\bibfnamefont {G.}~\bibnamefont {Spagnolli}}, \bibinfo {author}
  {\bibfnamefont {A.}~\bibnamefont {Trenkwalder}}, \bibinfo {author}
  {\bibfnamefont {M.}~\bibnamefont {Fattori}}, \bibinfo {author} {\bibfnamefont
  {M.}~\bibnamefont {Inguscio}},\ and\ \bibinfo {author} {\bibfnamefont
  {G.}~\bibnamefont {Modugno}},\ }\href@noop {} {\bibfield  {journal} {\bibinfo
   {journal} {Nature Physics}\ }\textbf {\bibinfo {volume} {11}},\ \bibinfo
  {pages} {554} (\bibinfo {year} {2015})}\BibitemShut {NoStop}%
\bibitem [{\citenamefont {Economou}\ and\ \citenamefont
  {Cohen}(1972)}]{PhysRevB.5.2931}%
  \BibitemOpen
  \bibfield  {author} {\bibinfo {author} {\bibfnamefont {E.~N.}\ \bibnamefont
  {Economou}}\ and\ \bibinfo {author} {\bibfnamefont {M.~H.}\ \bibnamefont
  {Cohen}},\ }\href {https://doi.org/10.1103/PhysRevB.5.2931} {\bibfield
  {journal} {\bibinfo  {journal} {Phys. Rev. B}\ }\textbf {\bibinfo {volume}
  {5}},\ \bibinfo {pages} {2931} (\bibinfo {year} {1972})}\BibitemShut
  {NoStop}%
\bibitem [{\citenamefont {Markos}(2006)}]{markos2006numerical}%
  \BibitemOpen
  \bibfield  {author} {\bibinfo {author} {\bibfnamefont {P.}~\bibnamefont
  {Markos}},\ }\href@noop {} {\bibfield  {journal} {\bibinfo  {journal} {arXiv
  preprint cond-mat/0609580}\ } (\bibinfo {year} {2006})}\BibitemShut {NoStop}%
\bibitem [{\citenamefont {Markoš}\ and\ \citenamefont
  {Kramer}(1993)}]{doi:10.1080/13642819308215292}%
  \BibitemOpen
  \bibfield  {author} {\bibinfo {author} {\bibfnamefont {P.}~\bibnamefont
  {Markoš}}\ and\ \bibinfo {author} {\bibfnamefont {B.}~\bibnamefont
  {Kramer}},\ }\href {https://doi.org/10.1080/13642819308215292} {\bibfield
  {journal} {\bibinfo  {journal} {Philosophical Magazine B}\ }\textbf {\bibinfo
  {volume} {68}},\ \bibinfo {pages} {357} (\bibinfo {year} {1993})},\ \Eprint
  {https://arxiv.org/abs/https://doi.org/10.1080/13642819308215292}
  {https://doi.org/10.1080/13642819308215292} \BibitemShut {NoStop}%
\bibitem [{\citenamefont {Evers}\ and\ \citenamefont
  {Mirlin}(2008)}]{RevModPhys.80.1355}%
  \BibitemOpen
  \bibfield  {author} {\bibinfo {author} {\bibfnamefont {F.}~\bibnamefont
  {Evers}}\ and\ \bibinfo {author} {\bibfnamefont {A.~D.}\ \bibnamefont
  {Mirlin}},\ }\href {https://doi.org/10.1103/RevModPhys.80.1355} {\bibfield
  {journal} {\bibinfo  {journal} {Rev. Mod. Phys.}\ }\textbf {\bibinfo {volume}
  {80}},\ \bibinfo {pages} {1355} (\bibinfo {year} {2008})}\BibitemShut
  {NoStop}%
\bibitem [{\citenamefont {Izrailev}\ and\ \citenamefont
  {Krokhin}(1999)}]{PhysRevLett.82.4062}%
  \BibitemOpen
  \bibfield  {author} {\bibinfo {author} {\bibfnamefont {F.~M.}\ \bibnamefont
  {Izrailev}}\ and\ \bibinfo {author} {\bibfnamefont {A.~A.}\ \bibnamefont
  {Krokhin}},\ }\href {https://doi.org/10.1103/PhysRevLett.82.4062} {\bibfield
  {journal} {\bibinfo  {journal} {Phys. Rev. Lett.}\ }\textbf {\bibinfo
  {volume} {82}},\ \bibinfo {pages} {4062} (\bibinfo {year}
  {1999})}\BibitemShut {NoStop}%
\bibitem [{\citenamefont {Mirlin}\ \emph {et~al.}(1996)\citenamefont {Mirlin},
  \citenamefont {Fyodorov}, \citenamefont {Dittes}, \citenamefont {Quezada},\
  and\ \citenamefont {Seligman}}]{mirlin1996transition}%
  \BibitemOpen
  \bibfield  {author} {\bibinfo {author} {\bibfnamefont {A.~D.}\ \bibnamefont
  {Mirlin}}, \bibinfo {author} {\bibfnamefont {Y.~V.}\ \bibnamefont
  {Fyodorov}}, \bibinfo {author} {\bibfnamefont {F.-M.}\ \bibnamefont
  {Dittes}}, \bibinfo {author} {\bibfnamefont {J.}~\bibnamefont {Quezada}},\
  and\ \bibinfo {author} {\bibfnamefont {T.~H.}\ \bibnamefont {Seligman}},\
  }\href@noop {} {\bibfield  {journal} {\bibinfo  {journal} {Physical Review
  E}\ }\textbf {\bibinfo {volume} {54}},\ \bibinfo {pages} {3221} (\bibinfo
  {year} {1996})}\BibitemShut {NoStop}%
\bibitem [{\citenamefont {Skinner}(1994)}]{doi:10.1021/j100061a002}%
  \BibitemOpen
  \bibfield  {author} {\bibinfo {author} {\bibfnamefont {J.~L.}\ \bibnamefont
  {Skinner}},\ }\href {https://doi.org/10.1021/j100061a002} {\bibfield
  {journal} {\bibinfo  {journal} {The Journal of Physical Chemistry}\ }\textbf
  {\bibinfo {volume} {98}},\ \bibinfo {pages} {2503} (\bibinfo {year}
  {1994})},\ \Eprint
  {https://arxiv.org/abs/https://doi.org/10.1021/j100061a002}
  {https://doi.org/10.1021/j100061a002} \BibitemShut {NoStop}%
\bibitem [{\citenamefont {Tian}(2019)}]{doi:10.1021/acsnano.9b02399}%
  \BibitemOpen
  \bibfield  {author} {\bibinfo {author} {\bibfnamefont {Z.}~\bibnamefont
  {Tian}},\ }\href {https://doi.org/10.1021/acsnano.9b02399} {\bibfield
  {journal} {\bibinfo  {journal} {ACS Nano}\ }\textbf {\bibinfo {volume}
  {13}},\ \bibinfo {pages} {3750} (\bibinfo {year} {2019})},\ \Eprint
  {https://arxiv.org/abs/https://doi.org/10.1021/acsnano.9b02399}
  {https://doi.org/10.1021/acsnano.9b02399} \BibitemShut {NoStop}%
\bibitem [{\citenamefont {Farka}\ \emph {et~al.}(2017)\citenamefont {Farka},
  \citenamefont {Coskun}, \citenamefont {Gasiorowski}, \citenamefont {Cobet},
  \citenamefont {Hingerl}, \citenamefont {Uiberlacker}, \citenamefont {Hild},
  \citenamefont {Greunz}, \citenamefont {Stifter}, \citenamefont {Sariciftci},
  \citenamefont {Menon}, \citenamefont {Schoefberger}, \citenamefont {Mardare},
  \citenamefont {Hassel}, \citenamefont {Schwarzinger}, \citenamefont
  {Scharber},\ and\ \citenamefont
  {Stadler}}]{https://doi.org/10.1002/aelm.201700050}%
  \BibitemOpen
  \bibfield  {author} {\bibinfo {author} {\bibfnamefont {D.}~\bibnamefont
  {Farka}}, \bibinfo {author} {\bibfnamefont {H.}~\bibnamefont {Coskun}},
  \bibinfo {author} {\bibfnamefont {J.}~\bibnamefont {Gasiorowski}}, \bibinfo
  {author} {\bibfnamefont {C.}~\bibnamefont {Cobet}}, \bibinfo {author}
  {\bibfnamefont {K.}~\bibnamefont {Hingerl}}, \bibinfo {author} {\bibfnamefont
  {L.~M.}\ \bibnamefont {Uiberlacker}}, \bibinfo {author} {\bibfnamefont
  {S.}~\bibnamefont {Hild}}, \bibinfo {author} {\bibfnamefont {T.}~\bibnamefont
  {Greunz}}, \bibinfo {author} {\bibfnamefont {D.}~\bibnamefont {Stifter}},
  \bibinfo {author} {\bibfnamefont {N.~S.}\ \bibnamefont {Sariciftci}},
  \bibinfo {author} {\bibfnamefont {R.}~\bibnamefont {Menon}}, \bibinfo
  {author} {\bibfnamefont {W.}~\bibnamefont {Schoefberger}}, \bibinfo {author}
  {\bibfnamefont {C.~C.}\ \bibnamefont {Mardare}}, \bibinfo {author}
  {\bibfnamefont {A.~W.}\ \bibnamefont {Hassel}}, \bibinfo {author}
  {\bibfnamefont {C.}~\bibnamefont {Schwarzinger}}, \bibinfo {author}
  {\bibfnamefont {M.~C.}\ \bibnamefont {Scharber}},\ and\ \bibinfo {author}
  {\bibfnamefont {P.}~\bibnamefont {Stadler}},\ }\href
  {https://doi.org/https://doi.org/10.1002/aelm.201700050} {\bibfield
  {journal} {\bibinfo  {journal} {Advanced Electronic Materials}\ }\textbf
  {\bibinfo {volume} {3}},\ \bibinfo {pages} {1700050} (\bibinfo {year}
  {2017})},\ \Eprint
  {https://arxiv.org/abs/https://onlinelibrary.wiley.com/doi/pdf/10.1002/aelm.201700050}
  {https://onlinelibrary.wiley.com/doi/pdf/10.1002/aelm.201700050} \BibitemShut
  {NoStop}%
\bibitem [{\citenamefont {Chen}\ \emph {et~al.}(2014)\citenamefont {Chen},
  \citenamefont {Roushan}, \citenamefont {Sank}, \citenamefont {Neill},
  \citenamefont {Lucero}, \citenamefont {Mariantoni}, \citenamefont {Barends},
  \citenamefont {Chiaro}, \citenamefont {Kelly}, \citenamefont {Megrant},
  \citenamefont {Mutus}, \citenamefont {O'Malley}, \citenamefont {Vainsencher},
  \citenamefont {Wenner}, \citenamefont {White}, \citenamefont {Yin},
  \citenamefont {Cleland},\ and\ \citenamefont {Martinis}}]{Chen2014}%
  \BibitemOpen
  \bibfield  {author} {\bibinfo {author} {\bibfnamefont {Y.}~\bibnamefont
  {Chen}}, \bibinfo {author} {\bibfnamefont {P.}~\bibnamefont {Roushan}},
  \bibinfo {author} {\bibfnamefont {D.}~\bibnamefont {Sank}}, \bibinfo {author}
  {\bibfnamefont {C.}~\bibnamefont {Neill}}, \bibinfo {author} {\bibfnamefont
  {E.}~\bibnamefont {Lucero}}, \bibinfo {author} {\bibfnamefont
  {M.}~\bibnamefont {Mariantoni}}, \bibinfo {author} {\bibfnamefont
  {R.}~\bibnamefont {Barends}}, \bibinfo {author} {\bibfnamefont
  {B.}~\bibnamefont {Chiaro}}, \bibinfo {author} {\bibfnamefont
  {J.}~\bibnamefont {Kelly}}, \bibinfo {author} {\bibfnamefont
  {A.}~\bibnamefont {Megrant}}, \bibinfo {author} {\bibfnamefont {J.~Y.}\
  \bibnamefont {Mutus}}, \bibinfo {author} {\bibfnamefont {P.~J.~J.}\
  \bibnamefont {O'Malley}}, \bibinfo {author} {\bibfnamefont {A.}~\bibnamefont
  {Vainsencher}}, \bibinfo {author} {\bibfnamefont {J.}~\bibnamefont {Wenner}},
  \bibinfo {author} {\bibfnamefont {T.~C.}\ \bibnamefont {White}}, \bibinfo
  {author} {\bibfnamefont {Y.}~\bibnamefont {Yin}}, \bibinfo {author}
  {\bibfnamefont {A.~N.}\ \bibnamefont {Cleland}},\ and\ \bibinfo {author}
  {\bibfnamefont {J.~M.}\ \bibnamefont {Martinis}},\ }\href
  {https://doi.org/10.1038/ncomms6184} {\bibfield  {journal} {\bibinfo
  {journal} {Nature Communications}\ }\textbf {\bibinfo {volume} {5}},\
  \bibinfo {pages} {5184} (\bibinfo {year} {2014})}\BibitemShut {NoStop}%
\bibitem [{\citenamefont {Poduval}\ and\ \citenamefont
  {Das~Sarma}(2023)}]{PhysRevB.107.174204}%
  \BibitemOpen
  \bibfield  {author} {\bibinfo {author} {\bibfnamefont {P.~P.}\ \bibnamefont
  {Poduval}}\ and\ \bibinfo {author} {\bibfnamefont {S.}~\bibnamefont
  {Das~Sarma}},\ }\href {https://doi.org/10.1103/PhysRevB.107.174204}
  {\bibfield  {journal} {\bibinfo  {journal} {Phys. Rev. B}\ }\textbf {\bibinfo
  {volume} {107}},\ \bibinfo {pages} {174204} (\bibinfo {year}
  {2023})}\BibitemShut {NoStop}%
\bibitem [{\citenamefont {Holthaus}\ \emph {et~al.}(1995)\citenamefont
  {Holthaus}, \citenamefont {Ristow},\ and\ \citenamefont
  {Hone}}]{PhysRevLett.75.3914}%
  \BibitemOpen
  \bibfield  {author} {\bibinfo {author} {\bibfnamefont {M.}~\bibnamefont
  {Holthaus}}, \bibinfo {author} {\bibfnamefont {G.~H.}\ \bibnamefont
  {Ristow}},\ and\ \bibinfo {author} {\bibfnamefont {D.~W.}\ \bibnamefont
  {Hone}},\ }\href {https://doi.org/10.1103/PhysRevLett.75.3914} {\bibfield
  {journal} {\bibinfo  {journal} {Phys. Rev. Lett.}\ }\textbf {\bibinfo
  {volume} {75}},\ \bibinfo {pages} {3914} (\bibinfo {year}
  {1995})}\BibitemShut {NoStop}%
\bibitem [{\citenamefont {Bragan\ifmmode~\mbox{\c{c}}\else \c{c}\fi{}a}\ \emph
  {et~al.}(2015)\citenamefont {Bragan\ifmmode~\mbox{\c{c}}\else \c{c}\fi{}a},
  \citenamefont {Aguiar}, \citenamefont {Vu\ifmmode \check{c}\else
  \v{c}\fi{}i\ifmmode \check{c}\else \v{c}\fi{}evi\ifmmode~\acute{c}\else
  \'{c}\fi{}}, \citenamefont {Tanaskovi\ifmmode~\acute{c}\else \'{c}\fi{}},\
  and\ \citenamefont {Dobrosavljevi\ifmmode~\acute{c}\else
  \'{c}\fi{}}}]{PhysRevB.92.125143}%
  \BibitemOpen
  \bibfield  {author} {\bibinfo {author} {\bibfnamefont {H.}~\bibnamefont
  {Bragan\ifmmode~\mbox{\c{c}}\else \c{c}\fi{}a}}, \bibinfo {author}
  {\bibfnamefont {M.~C.~O.}\ \bibnamefont {Aguiar}}, \bibinfo {author}
  {\bibfnamefont {J.}~\bibnamefont {Vu\ifmmode \check{c}\else
  \v{c}\fi{}i\ifmmode \check{c}\else \v{c}\fi{}evi\ifmmode~\acute{c}\else
  \'{c}\fi{}}}, \bibinfo {author} {\bibfnamefont {D.}~\bibnamefont
  {Tanaskovi\ifmmode~\acute{c}\else \'{c}\fi{}}},\ and\ \bibinfo {author}
  {\bibfnamefont {V.}~\bibnamefont {Dobrosavljevi\ifmmode~\acute{c}\else
  \'{c}\fi{}}},\ }\href {https://doi.org/10.1103/PhysRevB.92.125143} {\bibfield
   {journal} {\bibinfo  {journal} {Phys. Rev. B}\ }\textbf {\bibinfo {volume}
  {92}},\ \bibinfo {pages} {125143} (\bibinfo {year} {2015})}\BibitemShut
  {NoStop}%
\bibitem [{\citenamefont {Liao}\ \emph {et~al.}(2015)\citenamefont {Liao},
  \citenamefont {Ou}, \citenamefont {Feng}, \citenamefont {Yang}, \citenamefont
  {Lin}, \citenamefont {Yang}, \citenamefont {Wu}, \citenamefont {He},
  \citenamefont {Ma}, \citenamefont {Xue},\ and\ \citenamefont
  {Li}}]{PhysRevLett.114.216601}%
  \BibitemOpen
  \bibfield  {author} {\bibinfo {author} {\bibfnamefont {J.}~\bibnamefont
  {Liao}}, \bibinfo {author} {\bibfnamefont {Y.}~\bibnamefont {Ou}}, \bibinfo
  {author} {\bibfnamefont {X.}~\bibnamefont {Feng}}, \bibinfo {author}
  {\bibfnamefont {S.}~\bibnamefont {Yang}}, \bibinfo {author} {\bibfnamefont
  {C.}~\bibnamefont {Lin}}, \bibinfo {author} {\bibfnamefont {W.}~\bibnamefont
  {Yang}}, \bibinfo {author} {\bibfnamefont {K.}~\bibnamefont {Wu}}, \bibinfo
  {author} {\bibfnamefont {K.}~\bibnamefont {He}}, \bibinfo {author}
  {\bibfnamefont {X.}~\bibnamefont {Ma}}, \bibinfo {author} {\bibfnamefont
  {Q.-K.}\ \bibnamefont {Xue}},\ and\ \bibinfo {author} {\bibfnamefont
  {Y.}~\bibnamefont {Li}},\ }\href
  {https://doi.org/10.1103/PhysRevLett.114.216601} {\bibfield  {journal}
  {\bibinfo  {journal} {Phys. Rev. Lett.}\ }\textbf {\bibinfo {volume} {114}},\
  \bibinfo {pages} {216601} (\bibinfo {year} {2015})}\BibitemShut {NoStop}%
\bibitem [{\citenamefont {Byczuk}\ \emph {et~al.}(2005)\citenamefont {Byczuk},
  \citenamefont {Hofstetter},\ and\ \citenamefont
  {Vollhardt}}]{PhysRevLett.94.056404}%
  \BibitemOpen
  \bibfield  {author} {\bibinfo {author} {\bibfnamefont {K.}~\bibnamefont
  {Byczuk}}, \bibinfo {author} {\bibfnamefont {W.}~\bibnamefont {Hofstetter}},\
  and\ \bibinfo {author} {\bibfnamefont {D.}~\bibnamefont {Vollhardt}},\ }\href
  {https://doi.org/10.1103/PhysRevLett.94.056404} {\bibfield  {journal}
  {\bibinfo  {journal} {Phys. Rev. Lett.}\ }\textbf {\bibinfo {volume} {94}},\
  \bibinfo {pages} {056404} (\bibinfo {year} {2005})}\BibitemShut {NoStop}%
\bibitem [{\citenamefont {Castro}\ \emph {et~al.}(2015)\citenamefont {Castro},
  \citenamefont {L\'opez-Sancho},\ and\ \citenamefont
  {Vozmediano}}]{PhysRevB.92.085410}%
  \BibitemOpen
  \bibfield  {author} {\bibinfo {author} {\bibfnamefont {E.~V.}\ \bibnamefont
  {Castro}}, \bibinfo {author} {\bibfnamefont {M.~P.}\ \bibnamefont
  {L\'opez-Sancho}},\ and\ \bibinfo {author} {\bibfnamefont {M.~A.~H.}\
  \bibnamefont {Vozmediano}},\ }\href
  {https://doi.org/10.1103/PhysRevB.92.085410} {\bibfield  {journal} {\bibinfo
  {journal} {Phys. Rev. B}\ }\textbf {\bibinfo {volume} {92}},\ \bibinfo
  {pages} {085410} (\bibinfo {year} {2015})}\BibitemShut {NoStop}%
\bibitem [{\citenamefont {Khomchenko}\ \emph {et~al.}(2023)\citenamefont
  {Khomchenko}, \citenamefont {Ouerdane},\ and\ \citenamefont
  {Benenti}}]{khomchenko2023influence}%
  \BibitemOpen
  \bibfield  {author} {\bibinfo {author} {\bibfnamefont {I.}~\bibnamefont
  {Khomchenko}}, \bibinfo {author} {\bibfnamefont {H.}~\bibnamefont
  {Ouerdane}},\ and\ \bibinfo {author} {\bibfnamefont {G.}~\bibnamefont
  {Benenti}},\ }\href@noop {} {\bibinfo {title} {Influence of the anderson
  transition on thermoelectric energy conversion in disordered electronic
  systems}} (\bibinfo {year} {2023}),\ \Eprint
  {https://arxiv.org/abs/2307.00921} {arXiv:2307.00921 [cond-mat.mtrl-sci]}
  \BibitemShut {NoStop}%
\bibitem [{\citenamefont {Yan}\ \emph {et~al.}(2023)\citenamefont {Yan},
  \citenamefont {Li}, \citenamefont {Zhao}, \citenamefont {Wu}, \citenamefont
  {Liu}, \citenamefont {Li}, \citenamefont {Fu}, \citenamefont {Xiao},
  \citenamefont {Ma},\ and\ \citenamefont {Jia}}]{Yan:23}%
  \BibitemOpen
  \bibfield  {author} {\bibinfo {author} {\bibfnamefont {W.}~\bibnamefont
  {Yan}}, \bibinfo {author} {\bibfnamefont {Y.}~\bibnamefont {Li}}, \bibinfo
  {author} {\bibfnamefont {H.}~\bibnamefont {Zhao}}, \bibinfo {author}
  {\bibfnamefont {J.}~\bibnamefont {Wu}}, \bibinfo {author} {\bibfnamefont
  {W.}~\bibnamefont {Liu}}, \bibinfo {author} {\bibfnamefont {P.}~\bibnamefont
  {Li}}, \bibinfo {author} {\bibfnamefont {Y.}~\bibnamefont {Fu}}, \bibinfo
  {author} {\bibfnamefont {L.}~\bibnamefont {Xiao}}, \bibinfo {author}
  {\bibfnamefont {J.}~\bibnamefont {Ma}},\ and\ \bibinfo {author}
  {\bibfnamefont {S.}~\bibnamefont {Jia}},\ }\href
  {https://doi.org/10.1364/OPTCON.499768} {\bibfield  {journal} {\bibinfo
  {journal} {Opt. Continuum}\ }\textbf {\bibinfo {volume} {2}},\ \bibinfo
  {pages} {2116} (\bibinfo {year} {2023})}\BibitemShut {NoStop}%
\bibitem [{\citenamefont {Thakkar}(2023)}]{thakkar2023effect}%
  \BibitemOpen
  \bibfield  {author} {\bibinfo {author} {\bibfnamefont {A.}~\bibnamefont
  {Thakkar}},\ }\href@noop {} {\  (\bibinfo {year} {2023})}\BibitemShut
  {NoStop}%
\bibitem [{\citenamefont {Schwartz}\ \emph {et~al.}(2007)\citenamefont
  {Schwartz}, \citenamefont {Bartal}, \citenamefont {Fishman},\ and\
  \citenamefont {Segev}}]{schwartz2007transport}%
  \BibitemOpen
  \bibfield  {author} {\bibinfo {author} {\bibfnamefont {T.}~\bibnamefont
  {Schwartz}}, \bibinfo {author} {\bibfnamefont {G.}~\bibnamefont {Bartal}},
  \bibinfo {author} {\bibfnamefont {S.}~\bibnamefont {Fishman}},\ and\ \bibinfo
  {author} {\bibfnamefont {M.}~\bibnamefont {Segev}},\ }\href@noop {}
  {\bibfield  {journal} {\bibinfo  {journal} {Nature}\ }\textbf {\bibinfo
  {volume} {446}},\ \bibinfo {pages} {52} (\bibinfo {year} {2007})}\BibitemShut
  {NoStop}%
\bibitem [{\citenamefont {Lahini}\ \emph {et~al.}(2008)\citenamefont {Lahini},
  \citenamefont {Avidan}, \citenamefont {Pozzi}, \citenamefont {Sorel},
  \citenamefont {Morandotti}, \citenamefont {Christodoulides},\ and\
  \citenamefont {Silberberg}}]{lahini2008anderson}%
  \BibitemOpen
  \bibfield  {author} {\bibinfo {author} {\bibfnamefont {Y.}~\bibnamefont
  {Lahini}}, \bibinfo {author} {\bibfnamefont {A.}~\bibnamefont {Avidan}},
  \bibinfo {author} {\bibfnamefont {F.}~\bibnamefont {Pozzi}}, \bibinfo
  {author} {\bibfnamefont {M.}~\bibnamefont {Sorel}}, \bibinfo {author}
  {\bibfnamefont {R.}~\bibnamefont {Morandotti}}, \bibinfo {author}
  {\bibfnamefont {D.~N.}\ \bibnamefont {Christodoulides}},\ and\ \bibinfo
  {author} {\bibfnamefont {Y.}~\bibnamefont {Silberberg}},\ }\href@noop {}
  {\bibfield  {journal} {\bibinfo  {journal} {Physical Review Letters}\
  }\textbf {\bibinfo {volume} {100}},\ \bibinfo {pages} {013906} (\bibinfo
  {year} {2008})}\BibitemShut {NoStop}%
\bibitem [{\citenamefont {Xiong}\ and\ \citenamefont
  {Xiong}(2007)}]{PhysRevB.76.214204}%
  \BibitemOpen
  \bibfield  {author} {\bibinfo {author} {\bibfnamefont {S.-J.}\ \bibnamefont
  {Xiong}}\ and\ \bibinfo {author} {\bibfnamefont {Y.}~\bibnamefont {Xiong}},\
  }\href {https://doi.org/10.1103/PhysRevB.76.214204} {\bibfield  {journal}
  {\bibinfo  {journal} {Phys. Rev. B}\ }\textbf {\bibinfo {volume} {76}},\
  \bibinfo {pages} {214204} (\bibinfo {year} {2007})}\BibitemShut {NoStop}%
\bibitem [{\citenamefont {Islam}\ \emph {et~al.}(2013)\citenamefont {Islam},
  \citenamefont {Senko}, \citenamefont {Campbell}, \citenamefont {Korenblit},
  \citenamefont {Smith}, \citenamefont {Lee}, \citenamefont {Edwards},
  \citenamefont {Wang}, \citenamefont {Freericks},\ and\ \citenamefont
  {Monroe}}]{doi:10.1126/science.1232296}%
  \BibitemOpen
  \bibfield  {author} {\bibinfo {author} {\bibfnamefont {R.}~\bibnamefont
  {Islam}}, \bibinfo {author} {\bibfnamefont {C.}~\bibnamefont {Senko}},
  \bibinfo {author} {\bibfnamefont {W.~C.}\ \bibnamefont {Campbell}}, \bibinfo
  {author} {\bibfnamefont {S.}~\bibnamefont {Korenblit}}, \bibinfo {author}
  {\bibfnamefont {J.}~\bibnamefont {Smith}}, \bibinfo {author} {\bibfnamefont
  {A.}~\bibnamefont {Lee}}, \bibinfo {author} {\bibfnamefont {E.~E.}\
  \bibnamefont {Edwards}}, \bibinfo {author} {\bibfnamefont {C.-C.~J.}\
  \bibnamefont {Wang}}, \bibinfo {author} {\bibfnamefont {J.~K.}\ \bibnamefont
  {Freericks}},\ and\ \bibinfo {author} {\bibfnamefont {C.}~\bibnamefont
  {Monroe}},\ }\href {https://doi.org/10.1126/science.1232296} {\bibfield
  {journal} {\bibinfo  {journal} {Science}\ }\textbf {\bibinfo {volume}
  {340}},\ \bibinfo {pages} {583} (\bibinfo {year} {2013})},\ \Eprint
  {https://arxiv.org/abs/https://www.science.org/doi/pdf/10.1126/science.1232296}
  {https://www.science.org/doi/pdf/10.1126/science.1232296} \BibitemShut
  {NoStop}%
\bibitem [{\citenamefont {P{\'e}rez-Gonz{\'a}lez}\ \emph
  {et~al.}(2019)\citenamefont {P{\'e}rez-Gonz{\'a}lez}, \citenamefont {Bello},
  \citenamefont {G{\'o}mez-Le{\'o}n},\ and\ \citenamefont
  {Platero}}]{perez2019interplay}%
  \BibitemOpen
  \bibfield  {author} {\bibinfo {author} {\bibfnamefont {B.}~\bibnamefont
  {P{\'e}rez-Gonz{\'a}lez}}, \bibinfo {author} {\bibfnamefont {M.}~\bibnamefont
  {Bello}}, \bibinfo {author} {\bibfnamefont {{\'A}.}~\bibnamefont
  {G{\'o}mez-Le{\'o}n}},\ and\ \bibinfo {author} {\bibfnamefont
  {G.}~\bibnamefont {Platero}},\ }\href@noop {} {\bibfield  {journal} {\bibinfo
   {journal} {Physical Review B}\ }\textbf {\bibinfo {volume} {99}},\ \bibinfo
  {pages} {035146} (\bibinfo {year} {2019})}\BibitemShut {NoStop}%
\bibitem [{\citenamefont {Celardo}\ \emph {et~al.}(2016)\citenamefont
  {Celardo}, \citenamefont {Kaiser},\ and\ \citenamefont
  {Borgonovi}}]{celardo2016shielding}%
  \BibitemOpen
  \bibfield  {author} {\bibinfo {author} {\bibfnamefont {G.}~\bibnamefont
  {Celardo}}, \bibinfo {author} {\bibfnamefont {R.}~\bibnamefont {Kaiser}},\
  and\ \bibinfo {author} {\bibfnamefont {F.}~\bibnamefont {Borgonovi}},\
  }\href@noop {} {\bibfield  {journal} {\bibinfo  {journal} {Physical Review
  B}\ }\textbf {\bibinfo {volume} {94}},\ \bibinfo {pages} {144206} (\bibinfo
  {year} {2016})}\BibitemShut {NoStop}%
\bibitem [{\citenamefont {Cantin}\ \emph {et~al.}(2018)\citenamefont {Cantin},
  \citenamefont {Xu},\ and\ \citenamefont {Krems}}]{cantin2018effect}%
  \BibitemOpen
  \bibfield  {author} {\bibinfo {author} {\bibfnamefont {J.~T.}\ \bibnamefont
  {Cantin}}, \bibinfo {author} {\bibfnamefont {T.}~\bibnamefont {Xu}},\ and\
  \bibinfo {author} {\bibfnamefont {R.~V.}\ \bibnamefont {Krems}},\ }\href@noop
  {} {\bibfield  {journal} {\bibinfo  {journal} {Physical Review B}\ }\textbf
  {\bibinfo {volume} {98}},\ \bibinfo {pages} {014204} (\bibinfo {year}
  {2018})}\BibitemShut {NoStop}%
\bibitem [{\citenamefont {Chattaraj}\ and\ \citenamefont
  {Krems}(2016)}]{chattaraj2016effects}%
  \BibitemOpen
  \bibfield  {author} {\bibinfo {author} {\bibfnamefont {T.}~\bibnamefont
  {Chattaraj}}\ and\ \bibinfo {author} {\bibfnamefont {R.}~\bibnamefont
  {Krems}},\ }\href@noop {} {\bibfield  {journal} {\bibinfo  {journal}
  {Physical Review A}\ }\textbf {\bibinfo {volume} {94}},\ \bibinfo {pages}
  {023601} (\bibinfo {year} {2016})}\BibitemShut {NoStop}%
\bibitem [{\citenamefont {Nag}\ and\ \citenamefont {Garg}(2019)}]{nag2019many}%
  \BibitemOpen
  \bibfield  {author} {\bibinfo {author} {\bibfnamefont {S.}~\bibnamefont
  {Nag}}\ and\ \bibinfo {author} {\bibfnamefont {A.}~\bibnamefont {Garg}},\
  }\href@noop {} {\bibfield  {journal} {\bibinfo  {journal} {Physical Review
  B}\ }\textbf {\bibinfo {volume} {99}},\ \bibinfo {pages} {224203} (\bibinfo
  {year} {2019})}\BibitemShut {NoStop}%
\bibitem [{\citenamefont {Ch{\'a}vez}\ \emph {et~al.}(2021)\citenamefont
  {Ch{\'a}vez}, \citenamefont {Mattiotti}, \citenamefont
  {M{\'e}ndez-Berm{\'u}dez}, \citenamefont {Borgonovi},\ and\ \citenamefont
  {Celardo}}]{chavez2021disorder}%
  \BibitemOpen
  \bibfield  {author} {\bibinfo {author} {\bibfnamefont {N.~C.}\ \bibnamefont
  {Ch{\'a}vez}}, \bibinfo {author} {\bibfnamefont {F.}~\bibnamefont
  {Mattiotti}}, \bibinfo {author} {\bibfnamefont {J.}~\bibnamefont
  {M{\'e}ndez-Berm{\'u}dez}}, \bibinfo {author} {\bibfnamefont
  {F.}~\bibnamefont {Borgonovi}},\ and\ \bibinfo {author} {\bibfnamefont
  {G.~L.}\ \bibnamefont {Celardo}},\ }\href@noop {} {\bibfield  {journal}
  {\bibinfo  {journal} {Physical Review Letters}\ }\textbf {\bibinfo {volume}
  {126}},\ \bibinfo {pages} {153201} (\bibinfo {year} {2021})}\BibitemShut
  {NoStop}%
\bibitem [{\citenamefont {Klein}(1993)}]{klein1993localization}%
  \BibitemOpen
  \bibfield  {author} {\bibinfo {author} {\bibfnamefont {A.}~\bibnamefont
  {Klein}},\ }\href@noop {} {\bibfield  {journal} {\bibinfo  {journal} {Braz.
  J. Phys}\ }\textbf {\bibinfo {volume} {23}},\ \bibinfo {pages} {363}
  (\bibinfo {year} {1993})}\BibitemShut {NoStop}%
\bibitem [{\citenamefont {Murphy}\ \emph {et~al.}(2011)\citenamefont {Murphy},
  \citenamefont {Wortis},\ and\ \citenamefont
  {Atkinson}}]{murphy2011generalized}%
  \BibitemOpen
  \bibfield  {author} {\bibinfo {author} {\bibfnamefont {N.}~\bibnamefont
  {Murphy}}, \bibinfo {author} {\bibfnamefont {R.}~\bibnamefont {Wortis}},\
  and\ \bibinfo {author} {\bibfnamefont {W.}~\bibnamefont {Atkinson}},\
  }\href@noop {} {\bibfield  {journal} {\bibinfo  {journal} {Physical Review
  B}\ }\textbf {\bibinfo {volume} {83}},\ \bibinfo {pages} {184206} (\bibinfo
  {year} {2011})}\BibitemShut {NoStop}%
\bibitem [{\citenamefont {Pruisken}(1985)}]{pruisken1985participation}%
  \BibitemOpen
  \bibfield  {author} {\bibinfo {author} {\bibfnamefont {A.~M.}\ \bibnamefont
  {Pruisken}},\ }\href@noop {} {\bibfield  {journal} {\bibinfo  {journal}
  {Physical Review B}\ }\textbf {\bibinfo {volume} {31}},\ \bibinfo {pages}
  {416} (\bibinfo {year} {1985})}\BibitemShut {NoStop}%
\bibitem [{\citenamefont {Calixto}\ and\ \citenamefont
  {Romera}(2015)}]{calixto2015inverse}%
  \BibitemOpen
  \bibfield  {author} {\bibinfo {author} {\bibfnamefont {M.}~\bibnamefont
  {Calixto}}\ and\ \bibinfo {author} {\bibfnamefont {E.}~\bibnamefont
  {Romera}},\ }\href@noop {} {\bibfield  {journal} {\bibinfo  {journal}
  {Journal of Statistical Mechanics: Theory and Experiment}\ }\textbf {\bibinfo
  {volume} {2015}},\ \bibinfo {pages} {P06029} (\bibinfo {year}
  {2015})}\BibitemShut {NoStop}%
\bibitem [{\citenamefont {Shklovskii}\ \emph {et~al.}(1993)\citenamefont
  {Shklovskii}, \citenamefont {Shapiro}, \citenamefont {Sears}, \citenamefont
  {Lambrianides},\ and\ \citenamefont {Shore}}]{PhysRevB.47.11487}%
  \BibitemOpen
  \bibfield  {author} {\bibinfo {author} {\bibfnamefont {B.~I.}\ \bibnamefont
  {Shklovskii}}, \bibinfo {author} {\bibfnamefont {B.}~\bibnamefont {Shapiro}},
  \bibinfo {author} {\bibfnamefont {B.~R.}\ \bibnamefont {Sears}}, \bibinfo
  {author} {\bibfnamefont {P.}~\bibnamefont {Lambrianides}},\ and\ \bibinfo
  {author} {\bibfnamefont {H.~B.}\ \bibnamefont {Shore}},\ }\href
  {https://doi.org/10.1103/PhysRevB.47.11487} {\bibfield  {journal} {\bibinfo
  {journal} {Phys. Rev. B}\ }\textbf {\bibinfo {volume} {47}},\ \bibinfo
  {pages} {11487} (\bibinfo {year} {1993})}\BibitemShut {NoStop}%
\bibitem [{\citenamefont {Deng}\ \emph {et~al.}(2019)\citenamefont {Deng},
  \citenamefont {Ray}, \citenamefont {Sinha}, \citenamefont {Shlyapnikov},\
  and\ \citenamefont {Santos}}]{PhysRevLett.123.025301}%
  \BibitemOpen
  \bibfield  {author} {\bibinfo {author} {\bibfnamefont {X.}~\bibnamefont
  {Deng}}, \bibinfo {author} {\bibfnamefont {S.}~\bibnamefont {Ray}}, \bibinfo
  {author} {\bibfnamefont {S.}~\bibnamefont {Sinha}}, \bibinfo {author}
  {\bibfnamefont {G.~V.}\ \bibnamefont {Shlyapnikov}},\ and\ \bibinfo {author}
  {\bibfnamefont {L.}~\bibnamefont {Santos}},\ }\href
  {https://doi.org/10.1103/PhysRevLett.123.025301} {\bibfield  {journal}
  {\bibinfo  {journal} {Phys. Rev. Lett.}\ }\textbf {\bibinfo {volume} {123}},\
  \bibinfo {pages} {025301} (\bibinfo {year} {2019})}\BibitemShut {NoStop}%
\bibitem [{\citenamefont {Biroli}\ \emph {et~al.}(2012)\citenamefont {Biroli},
  \citenamefont {Ribeiro-Teixeira},\ and\ \citenamefont
  {Tarzia}}]{biroli2012differencelevelstatisticsergodicity}%
  \BibitemOpen
  \bibfield  {author} {\bibinfo {author} {\bibfnamefont {G.}~\bibnamefont
  {Biroli}}, \bibinfo {author} {\bibfnamefont {A.~C.}\ \bibnamefont
  {Ribeiro-Teixeira}},\ and\ \bibinfo {author} {\bibfnamefont {M.}~\bibnamefont
  {Tarzia}},\ }\href {https://arxiv.org/abs/1211.7334} {\bibinfo {title}
  {Difference between level statistics, ergodicity and localization transitions
  on the bethe lattice}} (\bibinfo {year} {2012}),\ \Eprint
  {https://arxiv.org/abs/1211.7334} {arXiv:1211.7334 [cond-mat.dis-nn]}
  \BibitemShut {NoStop}%
\bibitem [{\citenamefont {Roy}\ and\ \citenamefont
  {Sharma}(2018)}]{PhysRevB.97.125116}%
  \BibitemOpen
  \bibfield  {author} {\bibinfo {author} {\bibfnamefont {N.}~\bibnamefont
  {Roy}}\ and\ \bibinfo {author} {\bibfnamefont {A.}~\bibnamefont {Sharma}},\
  }\href {https://doi.org/10.1103/PhysRevB.97.125116} {\bibfield  {journal}
  {\bibinfo  {journal} {Phys. Rev. B}\ }\textbf {\bibinfo {volume} {97}},\
  \bibinfo {pages} {125116} (\bibinfo {year} {2018})}\BibitemShut {NoStop}%
\bibitem [{\citenamefont {Atas}\ \emph {et~al.}(2013)\citenamefont {Atas},
  \citenamefont {Bogomolny}, \citenamefont {Giraud},\ and\ \citenamefont
  {Roux}}]{PhysRevLett.110.084101}%
  \BibitemOpen
  \bibfield  {author} {\bibinfo {author} {\bibfnamefont {Y.~Y.}\ \bibnamefont
  {Atas}}, \bibinfo {author} {\bibfnamefont {E.}~\bibnamefont {Bogomolny}},
  \bibinfo {author} {\bibfnamefont {O.}~\bibnamefont {Giraud}},\ and\ \bibinfo
  {author} {\bibfnamefont {G.}~\bibnamefont {Roux}},\ }\href
  {https://doi.org/10.1103/PhysRevLett.110.084101} {\bibfield  {journal}
  {\bibinfo  {journal} {Phys. Rev. Lett.}\ }\textbf {\bibinfo {volume} {110}},\
  \bibinfo {pages} {084101} (\bibinfo {year} {2013})}\BibitemShut {NoStop}%
\bibitem [{\citenamefont {Oganesyan}\ and\ \citenamefont
  {Huse}(2007)}]{PhysRevB.75.155111}%
  \BibitemOpen
  \bibfield  {author} {\bibinfo {author} {\bibfnamefont {V.}~\bibnamefont
  {Oganesyan}}\ and\ \bibinfo {author} {\bibfnamefont {D.~A.}\ \bibnamefont
  {Huse}},\ }\href {https://doi.org/10.1103/PhysRevB.75.155111} {\bibfield
  {journal} {\bibinfo  {journal} {Phys. Rev. B}\ }\textbf {\bibinfo {volume}
  {75}},\ \bibinfo {pages} {155111} (\bibinfo {year} {2007})}\BibitemShut
  {NoStop}%
\bibitem [{\citenamefont {Einstein}\ \emph {et~al.}(1935)\citenamefont
  {Einstein}, \citenamefont {Podolsky},\ and\ \citenamefont
  {Rosen}}]{PhysRev.47.777}%
  \BibitemOpen
  \bibfield  {author} {\bibinfo {author} {\bibfnamefont {A.}~\bibnamefont
  {Einstein}}, \bibinfo {author} {\bibfnamefont {B.}~\bibnamefont {Podolsky}},\
  and\ \bibinfo {author} {\bibfnamefont {N.}~\bibnamefont {Rosen}},\ }\href
  {https://doi.org/10.1103/PhysRev.47.777} {\bibfield  {journal} {\bibinfo
  {journal} {Phys. Rev.}\ }\textbf {\bibinfo {volume} {47}},\ \bibinfo {pages}
  {777} (\bibinfo {year} {1935})}\BibitemShut {NoStop}%
\bibitem [{\citenamefont {Raimond}\ \emph {et~al.}(2001)\citenamefont
  {Raimond}, \citenamefont {Brune},\ and\ \citenamefont
  {Haroche}}]{RevModPhys.73.565}%
  \BibitemOpen
  \bibfield  {author} {\bibinfo {author} {\bibfnamefont {J.~M.}\ \bibnamefont
  {Raimond}}, \bibinfo {author} {\bibfnamefont {M.}~\bibnamefont {Brune}},\
  and\ \bibinfo {author} {\bibfnamefont {S.}~\bibnamefont {Haroche}},\ }\href
  {https://doi.org/10.1103/RevModPhys.73.565} {\bibfield  {journal} {\bibinfo
  {journal} {Rev. Mod. Phys.}\ }\textbf {\bibinfo {volume} {73}},\ \bibinfo
  {pages} {565} (\bibinfo {year} {2001})}\BibitemShut {NoStop}%
\bibitem [{\citenamefont {Schr\"{o}dinger}(1935)}]{schrodinger_1935}%
  \BibitemOpen
  \bibfield  {author} {\bibinfo {author} {\bibfnamefont {E.}~\bibnamefont
  {Schr\"{o}dinger}},\ }\href {https://doi.org/10.1017/S0305004100013554}
  {\bibfield  {journal} {\bibinfo  {journal} {Mathematical Proceedings of the
  Cambridge Philosophical Society}\ }\textbf {\bibinfo {volume} {31}},\
  \bibinfo {pages} {555–563} (\bibinfo {year} {1935})}\BibitemShut {NoStop}%
\bibitem [{\citenamefont {Osterloh}\ \emph {et~al.}(2002)\citenamefont
  {Osterloh}, \citenamefont {Amico}, \citenamefont {Falci},\ and\ \citenamefont
  {Fazio}}]{Osterloh2002}%
  \BibitemOpen
  \bibfield  {author} {\bibinfo {author} {\bibfnamefont {A.}~\bibnamefont
  {Osterloh}}, \bibinfo {author} {\bibfnamefont {L.}~\bibnamefont {Amico}},
  \bibinfo {author} {\bibfnamefont {G.}~\bibnamefont {Falci}},\ and\ \bibinfo
  {author} {\bibfnamefont {R.}~\bibnamefont {Fazio}},\ }\href
  {https://doi.org/10.1038/416608a} {\bibfield  {journal} {\bibinfo  {journal}
  {Nature}\ }\textbf {\bibinfo {volume} {416}},\ \bibinfo {pages} {608}
  (\bibinfo {year} {2002})}\BibitemShut {NoStop}%
\bibitem [{\citenamefont {Amico}\ \emph {et~al.}(2008)\citenamefont {Amico},
  \citenamefont {Fazio}, \citenamefont {Osterloh},\ and\ \citenamefont
  {Vedral}}]{RevModPhys.80.517}%
  \BibitemOpen
  \bibfield  {author} {\bibinfo {author} {\bibfnamefont {L.}~\bibnamefont
  {Amico}}, \bibinfo {author} {\bibfnamefont {R.}~\bibnamefont {Fazio}},
  \bibinfo {author} {\bibfnamefont {A.}~\bibnamefont {Osterloh}},\ and\
  \bibinfo {author} {\bibfnamefont {V.}~\bibnamefont {Vedral}},\ }\href
  {https://doi.org/10.1103/RevModPhys.80.517} {\bibfield  {journal} {\bibinfo
  {journal} {Rev. Mod. Phys.}\ }\textbf {\bibinfo {volume} {80}},\ \bibinfo
  {pages} {517} (\bibinfo {year} {2008})}\BibitemShut {NoStop}%
\bibitem [{\citenamefont {Horodecki}\ \emph {et~al.}(2009)\citenamefont
  {Horodecki}, \citenamefont {Horodecki}, \citenamefont {Horodecki},\ and\
  \citenamefont {Horodecki}}]{RevModPhys.81.865}%
  \BibitemOpen
  \bibfield  {author} {\bibinfo {author} {\bibfnamefont {R.}~\bibnamefont
  {Horodecki}}, \bibinfo {author} {\bibfnamefont {P.}~\bibnamefont
  {Horodecki}}, \bibinfo {author} {\bibfnamefont {M.}~\bibnamefont
  {Horodecki}},\ and\ \bibinfo {author} {\bibfnamefont {K.}~\bibnamefont
  {Horodecki}},\ }\href {https://doi.org/10.1103/RevModPhys.81.865} {\bibfield
  {journal} {\bibinfo  {journal} {Rev. Mod. Phys.}\ }\textbf {\bibinfo {volume}
  {81}},\ \bibinfo {pages} {865} (\bibinfo {year} {2009})}\BibitemShut
  {NoStop}%
\bibitem [{\citenamefont {Klich}(2006)}]{Klich_2006}%
  \BibitemOpen
  \bibfield  {author} {\bibinfo {author} {\bibfnamefont {I.}~\bibnamefont
  {Klich}},\ }\href {https://doi.org/10.1088/0305-4470/39/4/l02} {\bibfield
  {journal} {\bibinfo  {journal} {Journal of Physics A: Mathematical and
  General}\ }\textbf {\bibinfo {volume} {39}},\ \bibinfo {pages} {L85}
  (\bibinfo {year} {2006})}\BibitemShut {NoStop}%
\bibitem [{\citenamefont {Jia}\ \emph {et~al.}(2008)\citenamefont {Jia},
  \citenamefont {Subramaniam}, \citenamefont {Gruzberg},\ and\ \citenamefont
  {Chakravarty}}]{jia2008entanglement}%
  \BibitemOpen
  \bibfield  {author} {\bibinfo {author} {\bibfnamefont {X.}~\bibnamefont
  {Jia}}, \bibinfo {author} {\bibfnamefont {A.~R.}\ \bibnamefont
  {Subramaniam}}, \bibinfo {author} {\bibfnamefont {I.~A.}\ \bibnamefont
  {Gruzberg}},\ and\ \bibinfo {author} {\bibfnamefont {S.}~\bibnamefont
  {Chakravarty}},\ }\href@noop {} {\bibfield  {journal} {\bibinfo  {journal}
  {Physical Review B}\ }\textbf {\bibinfo {volume} {77}},\ \bibinfo {pages}
  {014208} (\bibinfo {year} {2008})}\BibitemShut {NoStop}%
\bibitem [{\citenamefont {Andrade}\ \emph {et~al.}(2014)\citenamefont
  {Andrade}, \citenamefont {Steudtner},\ and\ \citenamefont
  {Vojta}}]{andrade2014anderson}%
  \BibitemOpen
  \bibfield  {author} {\bibinfo {author} {\bibfnamefont {E.~C.}\ \bibnamefont
  {Andrade}}, \bibinfo {author} {\bibfnamefont {M.}~\bibnamefont {Steudtner}},\
  and\ \bibinfo {author} {\bibfnamefont {M.}~\bibnamefont {Vojta}},\
  }\href@noop {} {\bibfield  {journal} {\bibinfo  {journal} {Journal of
  Statistical Mechanics: Theory and Experiment}\ }\textbf {\bibinfo {volume}
  {2014}},\ \bibinfo {pages} {P07022} (\bibinfo {year} {2014})}\BibitemShut
  {NoStop}%
\bibitem [{\citenamefont {Laflorencie}(2022)}]{laflorencie2022entanglement}%
  \BibitemOpen
  \bibfield  {author} {\bibinfo {author} {\bibfnamefont {N.}~\bibnamefont
  {Laflorencie}},\ }in\ \href@noop {} {\emph {\bibinfo {booktitle}
  {Entanglement in Spin Chains: From Theory to Quantum Technology
  Applications}}}\ (\bibinfo  {publisher} {Springer},\ \bibinfo {year} {2022})\
  pp.\ \bibinfo {pages} {61--87}\BibitemShut {NoStop}%
\bibitem [{\citenamefont {Pastur}\ and\ \citenamefont
  {Slavin}(2014)}]{pastur2014area}%
  \BibitemOpen
  \bibfield  {author} {\bibinfo {author} {\bibfnamefont {L.}~\bibnamefont
  {Pastur}}\ and\ \bibinfo {author} {\bibfnamefont {V.}~\bibnamefont
  {Slavin}},\ }\href@noop {} {\bibfield  {journal} {\bibinfo  {journal}
  {Physical Review Letters}\ }\textbf {\bibinfo {volume} {113}},\ \bibinfo
  {pages} {150404} (\bibinfo {year} {2014})}\BibitemShut {NoStop}%
\bibitem [{\citenamefont {Pouranvari}\ and\ \citenamefont
  {Yang}(2014)}]{PhysRevB.89.115104}%
  \BibitemOpen
  \bibfield  {author} {\bibinfo {author} {\bibfnamefont {M.}~\bibnamefont
  {Pouranvari}}\ and\ \bibinfo {author} {\bibfnamefont {K.}~\bibnamefont
  {Yang}},\ }\href {https://doi.org/10.1103/PhysRevB.89.115104} {\bibfield
  {journal} {\bibinfo  {journal} {Phys. Rev. B}\ }\textbf {\bibinfo {volume}
  {89}},\ \bibinfo {pages} {115104} (\bibinfo {year} {2014})}\BibitemShut
  {NoStop}%
\bibitem [{\citenamefont {Pouranvari}(2020)}]{Pouranvari2020}%
  \BibitemOpen
  \bibfield  {author} {\bibinfo {author} {\bibfnamefont {M.}~\bibnamefont
  {Pouranvari}},\ }\href {https://doi.org/10.1140/epjb/e2020-10052-3}
  {\bibfield  {journal} {\bibinfo  {journal} {The European Physical Journal B}\
  }\textbf {\bibinfo {volume} {93}},\ \bibinfo {pages} {118} (\bibinfo {year}
  {2020})}\BibitemShut {NoStop}%
\bibitem [{\citenamefont {Pouranvari}(2023)}]{POURANVARI2023128908}%
  \BibitemOpen
  \bibfield  {author} {\bibinfo {author} {\bibfnamefont {M.}~\bibnamefont
  {Pouranvari}},\ }\href
  {https://doi.org/https://doi.org/10.1016/j.physa.2023.128908} {\bibfield
  {journal} {\bibinfo  {journal} {Physica A: Statistical Mechanics and its
  Applications}\ }\textbf {\bibinfo {volume} {624}},\ \bibinfo {pages} {128908}
  (\bibinfo {year} {2023})}\BibitemShut {NoStop}%
\bibitem [{\citenamefont {Pouranvari}\ and\ \citenamefont
  {Abouie}(2019)}]{PhysRevB.100.195109}%
  \BibitemOpen
  \bibfield  {author} {\bibinfo {author} {\bibfnamefont {M.}~\bibnamefont
  {Pouranvari}}\ and\ \bibinfo {author} {\bibfnamefont {J.}~\bibnamefont
  {Abouie}},\ }\href {https://doi.org/10.1103/PhysRevB.100.195109} {\bibfield
  {journal} {\bibinfo  {journal} {Phys. Rev. B}\ }\textbf {\bibinfo {volume}
  {100}},\ \bibinfo {pages} {195109} (\bibinfo {year} {2019})}\BibitemShut
  {NoStop}%
\bibitem [{\citenamefont {Pouranvari}\ \emph {et~al.}(2015)\citenamefont
  {Pouranvari}, \citenamefont {Zhang},\ and\ \citenamefont
  {Yang}}]{https://doi.org/10.1155/2015/397630}%
  \BibitemOpen
  \bibfield  {author} {\bibinfo {author} {\bibfnamefont {M.}~\bibnamefont
  {Pouranvari}}, \bibinfo {author} {\bibfnamefont {Y.}~\bibnamefont {Zhang}},\
  and\ \bibinfo {author} {\bibfnamefont {K.}~\bibnamefont {Yang}},\ }\href
  {https://doi.org/https://doi.org/10.1155/2015/397630} {\bibfield  {journal}
  {\bibinfo  {journal} {Advances in Condensed Matter Physics}\ }\textbf
  {\bibinfo {volume} {2015}},\ \bibinfo {pages} {397630} (\bibinfo {year}
  {2015})},\ \Eprint
  {https://arxiv.org/abs/https://onlinelibrary.wiley.com/doi/pdf/10.1155/2015/397630}
  {https://onlinelibrary.wiley.com/doi/pdf/10.1155/2015/397630} \BibitemShut
  {NoStop}%
\bibitem [{\citenamefont {Wolf}(2006)}]{wolf2006violation}%
  \BibitemOpen
  \bibfield  {author} {\bibinfo {author} {\bibfnamefont {M.~M.}\ \bibnamefont
  {Wolf}},\ }\href@noop {} {\bibfield  {journal} {\bibinfo  {journal} {Physical
  review letters}\ }\textbf {\bibinfo {volume} {96}},\ \bibinfo {pages}
  {010404} (\bibinfo {year} {2006})}\BibitemShut {NoStop}%
\bibitem [{\citenamefont {Gioev}\ and\ \citenamefont
  {Klich}(2006)}]{gioev2006entanglement}%
  \BibitemOpen
  \bibfield  {author} {\bibinfo {author} {\bibfnamefont {D.}~\bibnamefont
  {Gioev}}\ and\ \bibinfo {author} {\bibfnamefont {I.}~\bibnamefont {Klich}},\
  }\href@noop {} {\bibfield  {journal} {\bibinfo  {journal} {Physical review
  letters}\ }\textbf {\bibinfo {volume} {96}},\ \bibinfo {pages} {100503}
  (\bibinfo {year} {2006})}\BibitemShut {NoStop}%
\bibitem [{\citenamefont {Barthel}\ \emph {et~al.}(2006)\citenamefont
  {Barthel}, \citenamefont {Chung},\ and\ \citenamefont
  {Schollw\"ock}}]{PhysRevA.74.022329}%
  \BibitemOpen
  \bibfield  {author} {\bibinfo {author} {\bibfnamefont {T.}~\bibnamefont
  {Barthel}}, \bibinfo {author} {\bibfnamefont {M.-C.}\ \bibnamefont {Chung}},\
  and\ \bibinfo {author} {\bibfnamefont {U.}~\bibnamefont {Schollw\"ock}},\
  }\href {https://doi.org/10.1103/PhysRevA.74.022329} {\bibfield  {journal}
  {\bibinfo  {journal} {Phys. Rev. A}\ }\textbf {\bibinfo {volume} {74}},\
  \bibinfo {pages} {022329} (\bibinfo {year} {2006})}\BibitemShut {NoStop}%
\bibitem [{\citenamefont {Li}\ \emph {et~al.}(2006{\natexlab{a}})\citenamefont
  {Li}, \citenamefont {Ding}, \citenamefont {Yu}, \citenamefont {Roscilde},\
  and\ \citenamefont {Haas}}]{li2006scaling1}%
  \BibitemOpen
  \bibfield  {author} {\bibinfo {author} {\bibfnamefont {W.}~\bibnamefont
  {Li}}, \bibinfo {author} {\bibfnamefont {L.}~\bibnamefont {Ding}}, \bibinfo
  {author} {\bibfnamefont {R.}~\bibnamefont {Yu}}, \bibinfo {author}
  {\bibfnamefont {T.}~\bibnamefont {Roscilde}},\ and\ \bibinfo {author}
  {\bibfnamefont {S.}~\bibnamefont {Haas}},\ }\href@noop {} {\bibfield
  {journal} {\bibinfo  {journal} {Physical Review B—Condensed Matter and
  Materials Physics}\ }\textbf {\bibinfo {volume} {74}},\ \bibinfo {pages}
  {073103} (\bibinfo {year} {2006}{\natexlab{a}})}\BibitemShut {NoStop}%
\bibitem [{\citenamefont {Li}\ \emph {et~al.}(2006{\natexlab{b}})\citenamefont
  {Li}, \citenamefont {Ding}, \citenamefont {Yu}, \citenamefont {Roscilde},\
  and\ \citenamefont {Haas}}]{li2006scaling2}%
  \BibitemOpen
  \bibfield  {author} {\bibinfo {author} {\bibfnamefont {W.}~\bibnamefont
  {Li}}, \bibinfo {author} {\bibfnamefont {L.}~\bibnamefont {Ding}}, \bibinfo
  {author} {\bibfnamefont {R.}~\bibnamefont {Yu}}, \bibinfo {author}
  {\bibfnamefont {T.}~\bibnamefont {Roscilde}},\ and\ \bibinfo {author}
  {\bibfnamefont {S.}~\bibnamefont {Haas}},\ }\href@noop {} {\bibfield
  {journal} {\bibinfo  {journal} {Physical Review B—Condensed Matter and
  Materials Physics}\ }\textbf {\bibinfo {volume} {74}},\ \bibinfo {pages}
  {073103} (\bibinfo {year} {2006}{\natexlab{b}})}\BibitemShut {NoStop}%
\bibitem [{\citenamefont {Cramer}\ \emph {et~al.}(2007)\citenamefont {Cramer},
  \citenamefont {Eisert},\ and\ \citenamefont
  {Plenio}}]{PhysRevLett.98.220603}%
  \BibitemOpen
  \bibfield  {author} {\bibinfo {author} {\bibfnamefont {M.}~\bibnamefont
  {Cramer}}, \bibinfo {author} {\bibfnamefont {J.}~\bibnamefont {Eisert}},\
  and\ \bibinfo {author} {\bibfnamefont {M.~B.}\ \bibnamefont {Plenio}},\
  }\href {https://doi.org/10.1103/PhysRevLett.98.220603} {\bibfield  {journal}
  {\bibinfo  {journal} {Phys. Rev. Lett.}\ }\textbf {\bibinfo {volume} {98}},\
  \bibinfo {pages} {220603} (\bibinfo {year} {2007})}\BibitemShut {NoStop}%
\bibitem [{\citenamefont {Alet}\ and\ \citenamefont
  {Laflorencie}(2018)}]{alet2018many}%
  \BibitemOpen
  \bibfield  {author} {\bibinfo {author} {\bibfnamefont {F.}~\bibnamefont
  {Alet}}\ and\ \bibinfo {author} {\bibfnamefont {N.}~\bibnamefont
  {Laflorencie}},\ }\href@noop {} {\bibfield  {journal} {\bibinfo  {journal}
  {Comptes Rendus Physique}\ }\textbf {\bibinfo {volume} {19}},\ \bibinfo
  {pages} {498} (\bibinfo {year} {2018})}\BibitemShut {NoStop}%
\bibitem [{\citenamefont {Pal}\ and\ \citenamefont {Huse}(2010)}]{pal2010many}%
  \BibitemOpen
  \bibfield  {author} {\bibinfo {author} {\bibfnamefont {A.}~\bibnamefont
  {Pal}}\ and\ \bibinfo {author} {\bibfnamefont {D.~A.}\ \bibnamefont {Huse}},\
  }\href@noop {} {\bibfield  {journal} {\bibinfo  {journal} {Physical Review
  B—Condensed Matter and Materials Physics}\ }\textbf {\bibinfo {volume}
  {82}},\ \bibinfo {pages} {174411} (\bibinfo {year} {2010})}\BibitemShut
  {NoStop}%
\bibitem [{\citenamefont {Abanin}\ \emph {et~al.}(2019)\citenamefont {Abanin},
  \citenamefont {Altman}, \citenamefont {Bloch},\ and\ \citenamefont
  {Serbyn}}]{abanin2019colloquium}%
  \BibitemOpen
  \bibfield  {author} {\bibinfo {author} {\bibfnamefont {D.~A.}\ \bibnamefont
  {Abanin}}, \bibinfo {author} {\bibfnamefont {E.}~\bibnamefont {Altman}},
  \bibinfo {author} {\bibfnamefont {I.}~\bibnamefont {Bloch}},\ and\ \bibinfo
  {author} {\bibfnamefont {M.}~\bibnamefont {Serbyn}},\ }\href@noop {}
  {\bibfield  {journal} {\bibinfo  {journal} {Reviews of Modern Physics}\
  }\textbf {\bibinfo {volume} {91}},\ \bibinfo {pages} {021001} (\bibinfo
  {year} {2019})}\BibitemShut {NoStop}%
\bibitem [{\citenamefont {Iyer}\ \emph {et~al.}(2013)\citenamefont {Iyer},
  \citenamefont {Oganesyan}, \citenamefont {Refael},\ and\ \citenamefont
  {Huse}}]{iyer2013many}%
  \BibitemOpen
  \bibfield  {author} {\bibinfo {author} {\bibfnamefont {S.}~\bibnamefont
  {Iyer}}, \bibinfo {author} {\bibfnamefont {V.}~\bibnamefont {Oganesyan}},
  \bibinfo {author} {\bibfnamefont {G.}~\bibnamefont {Refael}},\ and\ \bibinfo
  {author} {\bibfnamefont {D.~A.}\ \bibnamefont {Huse}},\ }\href@noop {}
  {\bibfield  {journal} {\bibinfo  {journal} {Physical Review B—Condensed
  Matter and Materials Physics}\ }\textbf {\bibinfo {volume} {87}},\ \bibinfo
  {pages} {134202} (\bibinfo {year} {2013})}\BibitemShut {NoStop}%
\bibitem [{\citenamefont {Biddle}\ \emph {et~al.}(2009)\citenamefont {Biddle},
  \citenamefont {Wang}, \citenamefont {Priour~Jr},\ and\ \citenamefont
  {Das~Sarma}}]{biddle2009localization}%
  \BibitemOpen
  \bibfield  {author} {\bibinfo {author} {\bibfnamefont {J.}~\bibnamefont
  {Biddle}}, \bibinfo {author} {\bibfnamefont {B.}~\bibnamefont {Wang}},
  \bibinfo {author} {\bibfnamefont {D.}~\bibnamefont {Priour~Jr}},\ and\
  \bibinfo {author} {\bibfnamefont {S.}~\bibnamefont {Das~Sarma}},\ }\href@noop
  {} {\bibfield  {journal} {\bibinfo  {journal} {Physical Review A—Atomic,
  Molecular, and Optical Physics}\ }\textbf {\bibinfo {volume} {80}},\ \bibinfo
  {pages} {021603} (\bibinfo {year} {2009})}\BibitemShut {NoStop}%
\bibitem [{\citenamefont {An}\ \emph {et~al.}(2021)\citenamefont {An},
  \citenamefont {Padavi{\'c}}, \citenamefont {Meier}, \citenamefont {Hegde},
  \citenamefont {Ganeshan}, \citenamefont {Pixley}, \citenamefont
  {Vishveshwara},\ and\ \citenamefont {Gadway}}]{an2021interactions}%
  \BibitemOpen
  \bibfield  {author} {\bibinfo {author} {\bibfnamefont {F.~A.}\ \bibnamefont
  {An}}, \bibinfo {author} {\bibfnamefont {K.}~\bibnamefont {Padavi{\'c}}},
  \bibinfo {author} {\bibfnamefont {E.~J.}\ \bibnamefont {Meier}}, \bibinfo
  {author} {\bibfnamefont {S.}~\bibnamefont {Hegde}}, \bibinfo {author}
  {\bibfnamefont {S.}~\bibnamefont {Ganeshan}}, \bibinfo {author}
  {\bibfnamefont {J.}~\bibnamefont {Pixley}}, \bibinfo {author} {\bibfnamefont
  {S.}~\bibnamefont {Vishveshwara}},\ and\ \bibinfo {author} {\bibfnamefont
  {B.}~\bibnamefont {Gadway}},\ }\href@noop {} {\bibfield  {journal} {\bibinfo
  {journal} {Physical review letters}\ }\textbf {\bibinfo {volume} {126}},\
  \bibinfo {pages} {040603} (\bibinfo {year} {2021})}\BibitemShut {NoStop}%
\bibitem [{\citenamefont {Mastropietro}(2015)}]{mastropietro2015localization}%
  \BibitemOpen
  \bibfield  {author} {\bibinfo {author} {\bibfnamefont {V.}~\bibnamefont
  {Mastropietro}},\ }\href@noop {} {\bibfield  {journal} {\bibinfo  {journal}
  {Physical review letters}\ }\textbf {\bibinfo {volume} {115}},\ \bibinfo
  {pages} {180401} (\bibinfo {year} {2015})}\BibitemShut {NoStop}%
\bibitem [{\citenamefont {Ray}\ \emph {et~al.}(2018)\citenamefont {Ray},
  \citenamefont {Ghosh},\ and\ \citenamefont {Sinha}}]{ray2018drive}%
  \BibitemOpen
  \bibfield  {author} {\bibinfo {author} {\bibfnamefont {S.}~\bibnamefont
  {Ray}}, \bibinfo {author} {\bibfnamefont {A.}~\bibnamefont {Ghosh}},\ and\
  \bibinfo {author} {\bibfnamefont {S.}~\bibnamefont {Sinha}},\ }\href@noop {}
  {\bibfield  {journal} {\bibinfo  {journal} {Physical Review E}\ }\textbf
  {\bibinfo {volume} {97}},\ \bibinfo {pages} {010101} (\bibinfo {year}
  {2018})}\BibitemShut {NoStop}%
\bibitem [{\citenamefont {Bu}\ \emph {et~al.}(2022)\citenamefont {Bu},
  \citenamefont {Zhai},\ and\ \citenamefont {Yin}}]{bu2022quantum}%
  \BibitemOpen
  \bibfield  {author} {\bibinfo {author} {\bibfnamefont {X.}~\bibnamefont
  {Bu}}, \bibinfo {author} {\bibfnamefont {L.-J.}\ \bibnamefont {Zhai}},\ and\
  \bibinfo {author} {\bibfnamefont {S.}~\bibnamefont {Yin}},\ }\href@noop {}
  {\bibfield  {journal} {\bibinfo  {journal} {Physical Review B}\ }\textbf
  {\bibinfo {volume} {106}},\ \bibinfo {pages} {214208} (\bibinfo {year}
  {2022})}\BibitemShut {NoStop}%
\end{thebibliography}%
\end{document}